\documentclass[12pt]{article}
\title{Randomized selection
       revisited\thanks{Research supported by the State
       Committee for Scientific Research under Grant
       8T11A00622.}}
\author{Krzysztof C. Kiwiel\thanks{Systems Research Institute,
        Newelska 6, 01--447 Warsaw, Poland
        ({\tt kiwiel@ibspan.waw.pl})}}
\date{April 15, 2002}

\newcommand{\BbbF}{{\rm\normalcolor I\kern-.18em F}}
\newcommand{\BbbR}{{\rm\normalcolor I\kern-.18em R}}
\newcommand{\eqref}[1]{{\normalfont\normalcolor(\ref{#1})}}
\makeatletter
\def\proof{%
   \def\a##1{\begin{trivlist}\item[]{\bf\ignorespaces{##1}.}%
    \enspace\ignorespaces}%
   \def\b[##1]{\a{Proof\ \ignorespaces{##1}}}%
   \@ifnextchar[{\b}{\a{Proof}}}
\def\endproof{\end{trivlist}}
\def\qed{\relax\protect\ifmmode\ifinner\else\quad\fi\fi
    \hbox{\vbox{\hrule height.4pt\hbox{\vbox{\hrule height.4pt
    \hbox{\vrule width.4pt\vphantom{\normalsize A}\kern.5em
    \vrule width.4pt}\hrule height.4pt}}}}}
\newtoks\@stequation
\def\subequations{\refstepcounter{equation}%
\edef\@savedequation{\the\c@equation}%
\@stequation=\expandafter{\theequation}%   %only want \theequation
\edef\@savedtheequation{\the\@stequation}% %expanded once
\edef\oldtheequation{\theequation}%
\setcounter{equation}{0}%
\def\theequation{\oldtheequation\alph{equation}}}%
\def\endsubequations{%
\setcounter{equation}{\@savedequation}%
\@stequation=\expandafter{\@savedtheequation}%
\edef\theequation{\the\@stequation}\global\@ignoretrue}
\def\@begintheorem#1#2{\trivlist
    \item[\hskip \labelsep{\bfseries #1\ #2.}]\itshape}
\def\@opargbegintheorem#1#2#3{\trivlist
    \item[\hskip \labelsep{\bfseries #1\ #2\ (#3).}]\itshape}
\@addtoreset{equation}{section}% Makes \section reset `equation' counter.
\def\theequation{\thesection.\arabic{equation}}
\@addtoreset{figure}{section}

\@addtoreset{table}{section}

\let\@@eqnsel=\relax
\def\@tempa{%
    \stepcounter{equation}%
    \def\@currentlabel{\p@equation\theequation}%
    \global\@eqnswtrue\m@th
    \global\@eqcnt\z@
    \tabskip\mathindent
    \let\\=\@eqncr
    \setlength\abovedisplayskip{\topsep}%
    \ifvmode
      \addtolength\abovedisplayskip{\partopsep}%
    \fi
    \addtolength\abovedisplayskip{\parskip}%
    \setlength\belowdisplayskip{\abovedisplayskip}%
    \setlength\belowdisplayshortskip{\abovedisplayskip}%
    \setlength\abovedisplayshortskip{\abovedisplayskip}%
    $$\everycr{}\halign to\linewidth% $$
    \bgroup
      \hskip\@centering
      $\displaystyle\tabskip\z@skip{##}$\@eqnsel&%
      \global\@eqcnt\@ne \hskip \tw@\arraycolsep \hfil${##}$\hfil&%
      \global\@eqcnt\tw@ \hskip \tw@\arraycolsep
        $\displaystyle{##}$\hfil \tabskip\@centering&%
      \global\@eqcnt\thr@@
        \hb@xt@\z@\bgroup\hss##\egroup\tabskip\z@skip\cr}%
\def\@tempb{%
   \stepcounter{equation}%
   \def\@currentlabel{\p@equation\theequation}%
   \global\@eqnswtrue
   \m@th
   \global\@eqcnt\z@
   \tabskip\@centering
   \let\\\@eqncr
   $$\everycr{}\halign to\displaywidth\bgroup
       \hskip\@centering$\displaystyle\tabskip\z@skip{##}$\@eqnsel
      &\global\@eqcnt\@ne\hskip \tw@\arraycolsep \hfil${##}$\hfil
      &\global\@eqcnt\tw@ \hskip \tw@\arraycolsep
         $\displaystyle{##}$\hfil\tabskip\@centering
      &\global\@eqcnt\thr@@ \hb@xt@\z@\bgroup\hss##\egroup
         \tabskip\z@skip
      \cr
}
\ifx\eqnarray\@tempa%     If the fleqn document-class option is in effect
    \def\eqnarray{%
    \stepcounter{equation}%
    \def\@currentlabel{\p@equation\theequation}%
    \global\@eqnswtrue\m@th
    \global\@eqcnt\z@
    \tabskip\mathindent
    \let\\=\@eqncr
    \setlength\abovedisplayskip{\topsep}%
    \ifvmode
      \addtolength\abovedisplayskip{\partopsep}%
    \fi
    \addtolength\abovedisplayskip{\parskip}%
    \setlength\belowdisplayskip{\abovedisplayskip}%
    \setlength\belowdisplayshortskip{\abovedisplayskip}%
    \setlength\abovedisplayshortskip{\abovedisplayskip}%
    $$\everycr{}\halign to\linewidth% $$
    \bgroup
      \hskip\@centering
      $\displaystyle\tabskip\z@skip{##}$\@eqnsel&%
      \global\@eqcnt\@ne
      \@@eqnsel%            \@@eqnsel has replaced \hskip \tw@\arraycolsep!!!
      \hfil${{}##{}}$\hfil&%              as in fixup.sty but textstyle!!!
      \global\@eqcnt\tw@
      \@@eqnsel%           \@@eqnsel has replaced \hskip \tw@\arraycolsep!!!
        $\displaystyle{##}$\hfil \tabskip\@centering&%
      \global\@eqcnt\thr@@
        \hb@xt@\z@\bgroup\hss##\egroup\tabskip\z@skip\cr}%
\else\ifx\eqnarray\@tempb%       Else try the default eqnarray environment.
   \def\eqnarray{%
   \stepcounter{equation}%
   \def\@currentlabel{\p@equation\theequation}%
   \global\@eqnswtrue
   \m@th
   \global\@eqcnt\z@
   \tabskip\@centering
   \let\\\@eqncr
   $$\everycr{}\halign to\displaywidth\bgroup
       \hskip\@centering$\displaystyle\tabskip\z@skip{##}$\@eqnsel
      &\global\@eqcnt\@ne
      \@@eqnsel%           \@@eqnsel has replaced \hskip \tw@\arraycolsep!!!
      \hfil${{}##{}}$\hfil%              as in fixup.sty but textstyle!!!
      &\global\@eqcnt\tw@
      \@@eqnsel%           \@@eqnsel has replaced \hskip \tw@\arraycolsep!!!
         $\displaystyle{##}$\hfil\tabskip\@centering
      &\global\@eqcnt\thr@@ \hb@xt@\z@\bgroup\hss##\egroup
         \tabskip\z@skip
      \cr}
\else 
\fi\fi
\def\@tempa{}			% Free up TeX's memory
\def\@tempb{}
\@ifundefined{chapter}{%
  \renewenvironment{thebibliography}[1]
     {\section*{\refname
        \@mkboth{\MakeUppercase\refname}{\MakeUppercase\refname}}%
      \list{\@biblabel{\@arabic\c@enumiv}}%
           {\settowidth\labelwidth{\@biblabel{#1}}%
            \leftmargin\labelwidth
            \advance\leftmargin\labelsep
            \itemsep \z@                 % Suppresses vertical separation.
            \@openbib@code
            \usecounter{enumiv}%
            \let\p@enumiv\@empty
            \renewcommand\theenumiv{\@arabic\c@enumiv}}%
      \sloppy
      \clubpenalty4000
      \@clubpenalty \clubpenalty
      \widowpenalty4000%
      \sfcode`\.\@m}
     {\def\@noitemerr
       {\@latex@warning{Empty `thebibliography' environment}}%
      \endlist}}%
{\renewenvironment{thebibliography}[1]
     {\section*{\bibname
        \@mkboth{\MakeUppercase\bibname}{\MakeUppercase\bibname}}%
      \list{\@biblabel{\@arabic\c@enumiv}}%
           {\settowidth\labelwidth{\@biblabel{#1}}%
            \leftmargin\labelwidth
            \advance\leftmargin\labelsep
            \itemsep \z@                 % Suppresses vertical separation.
            \@openbib@code
            \usecounter{enumiv}%
            \let\p@enumiv\@empty
            \renewcommand\theenumiv{\@arabic\c@enumiv}}%
      \sloppy
      \clubpenalty4000
      \@clubpenalty \clubpenalty
      \widowpenalty4000%
      \sfcode`\.\@m}
     {\def\@noitemerr
       {\@latex@warning{Empty `thebibliography' environment}}%
      \endlist}}%
\newcommand{\Argmax}{{\operator@font Arg}\max}
\newcommand{\Argmin}{{\operator@font Arg}\min}
\newcommand{\argmax}{{\operator@font arg}\max}
\newcommand{\argmin}{{\operator@font arg}\min}
\newcommand{\Exp}{\mathord{\operator@font E}}
\newcommand{\Prob}{\mathord{\operator@font P}}
\newcommand{\var}{\mathop{\operator@font var}}
\makeatother
\newtheorem{theorem}{Theorem}[section]
\newtheorem{algorithm}[theorem]{Algorithm}

\newtheorem{corollary}[theorem]{Corollary}

\newtheorem{fact}[theorem]{Fact}
\newtheorem{lemma}[theorem]{Lemma}

\newtheorem{remark}[theorem]{Remark}
\newtheorem{remarks}[theorem]{Remarks}
\begin{document}           % End of preamble and beginning of text.

\maketitle                 % Produces the title.

\begin{abstract}
\noindent
We show that several versions of Floyd and Rivest's algorithm
{\sc Select} for finding the $k$th smallest of $n$ elements require at
most $n+\min\{k,n-k\}+o(n)$ comparisons on average and with high
probability.  This rectifies the analysis of Floyd and Rivest, and
extends it to the case of nondistinct elements.  Our computational
results confirm that {\sc Select} may be the best algorithm in practice.
\end{abstract}

\begin{quotation}
\noindent{\bf Key words.} Selection, medians, partitioning,
computational complexity.
\end{quotation}

%\begin{quotation}
%\noindent{\bf MSC Subject Classifications.} 68W20, 68W05, 68Q25
%\end{quotation}

%\begin{quotation}
%\noindent{\bf Abbreviated title:} Randomized selection.
%\end{quotation}

%   *** SECTION 1 ***
\section{Introduction}
\label{s:intro}
The {\em selection problem\/} is defined as follows: Given a set
$X:=\{x_j\}_{j=1}^n$ of $n$ elements, a total order $<$ on $X$,
and an integer $1\le k\le n$, find the {\em $k$th smallest\/}
element of $X$, i.e., an element $x$ of $X$ for which there are at
most $k-1$ elements $x_j<x$ and at least $k$ elements $x_j\le x$.
The {\em median\/} of $X$ is the $\lceil n/2\rceil$th smallest
element of $X$.  (Since we are {\em not\/} assuming that the elements
are distinct, $X$ may be regarded as a multiset).

Selection is one of the fundamental problems in computer science.
It is used in the solution of other basic problems such as sorting and
finding convex hulls.  Hence its literature is too vast to be reviewed
here; see, e.g., \cite{dohaulzw:lbs,dozw:sm,dozw:msr} and
\cite[\S5.3.3]{knu:acpIII2}.  We only stress that most references
employ a comparison model (in which a selection algorithm is charged
only for comparisons between pairs of elements), assuming that the
elements are distinct.  Then, in the worst case, selection needs at
least $(2+\epsilon)n$ comparisons \cite{dozw:msr}, whereas the
pioneering algorithm of \cite{blflprrita:tbs} makes at most $5.43n$, its
first improvement of \cite{scpapi:fm} needs $3n+o(n)$, and the most
recent improvement in \cite{dozw:sm} takes $2.95n+o(n)$.  Thus a gap of
almost $50\%$ still remains between the best lower and upper bounds in
the worst case.

The average case is better understood.  Specifically, for
$k\le\lceil n/2\rceil$, at least $n+k-2$ comparisons are necessary
\cite{cumu:acs}, \cite[Ex.\ 5.3.3--25]{knu:acpIII2}, whereas the best
upper bound is $n+k+O(n^{1/2}\ln^{1/2}n)$
\cite[Eq.\ (5.3.3.16)]{knu:acpIII2}.  Yet this bound holds for a hardly
implementable theoretical scheme \cite[Ex.\ 5.3.3--24]{knu:acpIII2},
whereas a similar frequently cited bound for the algorithm {\sc Select}
of \cite{flri:etb} doesn't have a full proof, as noted in
\cite[Ex.\ 5.3.3--24]{knu:acpIII2} and \cite{poriti:eds}.
Significantly worse bounds hold for the classical algorithm {\sc Find}
of \cite{hoa:fa65}, also known as quickselect, which partitions $X$ by
using the median of a random sample of size $s\ge1$.  In particular, for
$k=\lceil n/2\rceil$, the upper bound is $3.39n+o(n)$ for $s=1$
\cite[Ex.\ 5.2.2--32]{knu:acpIII2} and $2.75n+o(n)$ for $s=3$
\cite{gru:mvh,kimapr:ahf}, whereas for finding an element of random
rank, the average cost is $3n+o(n)$ for $s=1$, $2.5n+o(n)$ for $s=3$
\cite{kimapr:ahf}, and $2n+o(n)$ when $s\to\infty$, $s/n\to0$ as
$n\to\infty$ \cite{maro:oss}.  In practice {\sc Find} is most popular,
because the algorithms of \cite{blflprrita:tbs,scpapi:fm} are much
slower on the average \cite{mus:iss,val:iss}.  For the general case of
nondistinct elements, little is known in theory about these algorithms,
but again {\sc Find} performs well in practice \cite{val:iss}.

Our aim is to rekindle theoretical and practical interest in the
algorithm {\sc Select} of \cite[\S2.1]{flri:etb} (the versions of
\cite[\S2.3]{flri:etb} and \cite{flri:asf} will be addressed
elsewhere).  We show that {\sc Select} performs very well in both theory
and practice, even when equal elements occur.  To outline our
contributions in more detail, we recall that {\sc Select} operates as
follows.  Using a small random sample, two elements $u$ and $v$ almost
sure to be just below and above the $k$th are found.  The remaining
elements are compared with $u$ and $v$ to create a small selection
problem on the elements between $u$ and $v$ that is quickly solved
recursively.  By taking a random subset as the sample, this approach
does well against any input ordering, both on average and with high
probability.

First, we revise {\sc Select} slightly to simplify our analysis.  Then,
without assuming that the elements are distinct, we show that
{\sc Select} needs at most $n+\min\{k,n-k\}+O(n^{2/3}\ln^{1/3}n)$
comparisons on average; this agrees with the result of
\cite[\S2.2]{flri:etb} which is based on an unproven assumption
\cite[\S5]{poriti:eds}.  Similar upper bounds are established for
versions that choose sample sizes as in
\cite{flri:asf,meh:osa,rei:ppa} and \cite[\S3.3]{mora:ra}.  Thus the
average costs of these versions reach the lower bounds of $1.5n+o(n)$
for median selection and $1.25n+o(n)$ for selecting an element of random
rank (yet the original sample size of \cite[\S2.2]{flri:etb} has the
best lower order term in its cost).  We also prove that nonrecursive
versions of {\sc Select}, which employ other selection or sorting
algorithms for small subproblems, require at most
$n+\min\{k,n-k\}+o(n)$ comparisons with high probability (e.g.,
$1-4n^{-2\beta}$ for a user-specified $\beta>0$); this extends and
strengthens the results of \cite[Thm 2]{gesi:chb},
\cite[Thm 2]{meh:osa} and \cite[Thm 3.5]{mora:ra}.

Since theoretical bounds alone needn't convince practitioners (who
may worry about hidden constants, etc.), a serious effort was made to
design a competitive implementation of {\sc Select}.  Here, as with
{\sc Find} and quicksort \cite{sed:qek}, the partitioning efficiency
is crucial.  In contrast with the observation of
\cite[p.\ 169]{flri:etb} that ``partitioning $X$ about both $u$ and $v$
[is] an inherently inefficient operation'', we introduce several
{\em quintary\/} schemes which perform well in practice.  As a
byproduct, we give a modification of the {\em ternary\/} partitioning
scheme of \cite{bemc:esf,bese:fas} that obviates subscript range
checking.

Relative to {\sc Find}, {\sc Select} requires only small additional
stack space for recursion, because sampling without replacement can be
done in place.  Still, it might seem that random sampling needs too much
time for random number generation.  (Hence several popular
implementations of {\sc Find} don't sample randomly, assuming that the
input file is in random order, whereas others \cite{val:iss} invoke
random sampling only when slow progress occurs.)  Yet our computational
experience shows that sampling doesn't hurt even on random inputs, and
it helps a lot on more difficult inputs (in fact our interest in
{\sc Select} was sparked by the poor performance of the implementation
of \cite{flri:asf} on several inputs of \cite{val:iss}).  Most
importantly, even for examples with relatively low comparison costs,
{\sc Select} beats quite sophisticated implementations of {\sc Find} by
a wide margin, in both comparison counts and computing times.  To save
space, only selected results are reported, but our experience on many
other inputs was similar.  In particular, empirical estimates of the
constants hidden in our bounds were always quite small.  Further, the
performance of {\sc Select} is extremely stable across a variety of
inputs, even for small input sizes (cf.\ \S\ref{ss:result}).  A
theoretical explanation of these features will be undertaken elsewhere.
For now, our experience supports the claim of \cite[\S1]{flri:etb} that
``the algorithm presented here is probably the best practical choice''.

The paper is organized as follows.  A general version of {\sc Select} is
introduced in \S\ref{s:alg}, and its basic features are analyzed in
\S\ref{s:analysis}.  The average performance of {\sc Select} is studied
in \S\ref{s:analrec}.  High probability bounds for nonrecursive versions
are derived in \S\ref{s:analnonrec}.  Partitioning schemes are discussed
in \S\ref{s:ternquint}.  Finally, our computational results are reported
in \S\ref{s:exp}.
%The Appendix contains proofs of certain technical results.
%Finally, we have a conclusion section.

Our notation is fairly standard.  $|A|$ denotes the cardinality of a set
$A$.  In a given probability space, $\Prob$ is the probability measure,
and $\Exp$ is the mean-value operator.
%
%   *** SECTION 2 ***
\section{The algorithm {\sc Select}}
\label{s:alg}
In this section we describe a general version of {\sc Select} in terms
of two auxiliary functions $s(n)$ and $g(n)$ (the sample size and rank
gap), which will be chosen later.  We omit their arguments in general,
as no confusion can arise.

{\sc Select} picks a small random sample $S$ from $X$ and two pivots $u$
and $v$ from $S$ such that $u\le x_k^*\le v$ with high probability,
where $x_k^*$ is the $k$th smallest element of $X$.  Partitioning $X$
into elements less than $u$, between $u$ and $v$, greater than $v$, and
equal to $u$ or $v$, {\sc Select} either detects that $u$ or $v$ equals
$x_k^*$, or determines a subset $\hat X$ of $X$ and an integer $\hat k$
such that $x_k^*$ may be selected recursively as the $\hat k$th smallest
element of $\hat X$.

Below is a detailed description of the algorithm.
%
%   *** ALGORITHM 2.1 ***
\begin{algorithm}
\label{alg:selABCDE}
\rm
\hfil\newline\noindent{\bf {\sc Select}$(X,k)$}
(Selects the $k$th smallest element of $X$, with $1\le k\le n:=|X|$)
\medbreak\noindent{\bf Step 1} ({\em Initiation\/}).
If $n=1$, return $x_1$.
Choose the sample size $s\le n-1$ and gap $g>0$.
\medbreak\noindent{\bf Step 2} ({\em Sample selection\/}).
Pick randomly a sample $S:=\{y_1,\ldots,y_s\}$ from $X$.
\medbreak\noindent{\bf Step 3} ({\em Pivot selection\/}).
Set $i_u:=\max\{\lceil ks/n-g\rceil,1\}$,
$i_v:=\min\{\lceil ks/n+g\rceil,s\}$.
Let $u$ and $v$ be the $i_u$th and $i_v$th smallest elements of $S$,
found by using {\sc Select} recursively.
\medbreak\noindent{\bf Step 4} ({\em Partitioning\/}).
By comparing each element $x$ of $X$ to $u$ and $v$, partition $X$
into $A:=\{x\in X:x<u\}$, $B:=\{x\in X:x=u\}$, $C:=\{x\in X:u<x<v\}$,
$D:=\{x\in X:x=v\}$, $E:=\{x\in X:v<x\}$.  If $k<n/2$, $x$
is compared to $v$ first, and to $u$ only if $x<v$ and $u<v$.  If
$k\ge n/2$, the order of the comparisons is reversed.
\medbreak\noindent{\bf Step 5} ({\em Stopping test\/}).
If $|A|<k\le|A\cup B|$ then return $u$; else if
$|A\cup B\cup C|<k\le n-|E|$ then return $v$.
\medbreak\noindent{\bf Step 6} ({\em Reduction\/}).
If $k\le|A|$, set $\hat X:=A$ and $\hat k:=k$; else if $n-|E|<k$, set
$\hat X:=E$ and $\hat k:=k-n+|E|$; else set $\hat X:=C$ and
$\hat k:=k-|A\cup B|$.  Set $\hat n:=|\hat X|$.
\medbreak\noindent{\bf Step 7} ({\em Recursion\/}).
%Set $X:=\hat X$, $k:=\hat k$, $n:=|X|$ and go to Step 1.
Return {\sc Select}$(\hat X,\hat k)$.
\end{algorithm}

A few remarks on the algorithm are in order.
%
%   *** REMARKS 2.2 ***
\begin{remarks}
\label{r:selABCDE}
\rm
(a)
The correctness and finiteness of {\sc Select} stem by induction from
the following observations.  The returns of Steps 1 and 5 deliver the
desired element.  At Step 6, $\hat X$ and $\hat k$ are chosen so that
the $k$th smallest element of $X$ is the $\hat k$th smallest element
of $\hat X$, and $\hat n<n$ (since $u,v\not\in\hat X$).  Also $|S|<n$
for the recursive calls at Step 3.
\par(b)
When Step 5 returns $u$ (or $v$), {\sc Select} may also return
information about the positions of the elements of $X$ relative to $u$
(or $v$). For instance, if $X$ is stored as an array, its $k$ smallest
elements may be placed first via interchanges at Step 4
(cf.\ \S\ref{s:ternquint}).  Hence after Step 3 finds $u$, we may remove
from $S$ its first $i_u$ smallest elements before extracting $v$.
Further, Step 4 need only compare $u$ and $v$ with the elements of
$X\setminus S$.
\par(c)
The following elementary property is needed in \S\ref{s:analrec}.
Let $c_n$ denote the maximum number of comparisons taken by {\sc Select}
on any input of size $n$.  Since Step 3 makes at most $c_{s}+c_{s-i_u}$
comparisons with $s<n$, Step 4 needs at most $2(n-s)$, and Step 7 takes
at most $c_{\hat n}$ with $\hat n<n$, by induction $c_{n}<\infty$ for
all $n$.
\end{remarks}
%
%   *** SECTION 3 ***
\section{Preliminary analysis}
\label{s:analysis}
In this section we analyze general features of sampling used by
{\sc Select}.
%
%   *** SUBSECTION 3.1 ***
\subsection{Sampling deviations and expectation bounds}
\label{ss:sampleexp}
Our analysis hinges on the following bound on the tail of the
hypergeometric distribution established in \cite{hoe:pis} and
rederived shortly in \cite{chv:thd}.
%
%   *** FACT 3.1 ***
\begin{fact}
\label{f:balls}
Let\/ $s$ balls be chosen uniformly at random from a set of\/ $n$ balls,
of which\/ $r$ are red, and\/ $r'$ be the random variable representing
the number of red balls drawn.  Let\/ $p:=r/n$.  Then
\begin{equation}
\Prob\left[\,r'\ge ps+g\,\right]\le e^{-2g^2\!/s}\quad\forall g\ge0.
\label{Pexpg}
\end{equation}
\end{fact}

We shall also need a simple version of the (left) Chebyshev inequality
\cite[\S2.4.2]{kor:hpt}.
%
%   *** FACT 3.2 ***
\begin{fact}
\label{f:Ebound}
Let\/ $z$ be a nonnegative random variable such that\/
$\Prob[z\le\zeta]=1$ for some constant\/ $\zeta$.  Then\/
$\Exp z\le t+\zeta\Prob[z>t]$ %\le t+\zeta\Prob[z\ge t]$
for all nonnegative real numbers\/ $t$.
\end{fact}
%
%   *** SUBSECTION 3.2 ***
\subsection{Sample ranks and partitioning efficiency}
\label{ss:samplerank}
Denote by $x_1^*\le\ldots\le x_n^*$ and $y_1^*\le\ldots\le y_s^*$ the
sorted elements of the input set $X$ and the sample set $S$,
respectively.  Thus $x_k^*$ is the $k$th smallest element of $X$,
whereas $u=y_{i_u}^*$ and $v=y_{i_v}^*$ at Step 3.  This notation
facilitates showing that for the bounding indices
\begin{equation}
k_l:=\max\left\{\,\lceil k-2gn/s\rceil,1\,\right\}
\quad\mbox{and}\quad
k_r:=\min\left\{\,\lceil k+2gn/s\rceil,n\,\right\},
\label{klkr}
\end{equation}
we have $x_{k_l}^*\le u\le x_k^*\le v\le x_{k_r}^*$ with high
probability for suitable choices of $s$ and $g$.
%
%   *** LEMMA 3.3 ***
\begin{lemma}
\label{l:rank}
{\rm(a)}
$\Prob[x_k^*<u]\le e^{-2g^2\!/s}$ if\/ $i_u=\lceil ks/n-g\rceil$.
\par\indent\rlap{\rm(b)}\hphantom{\rm(a)}
$\Prob[u<x_{k_l}^*]\le e^{-2g^2\!/s}$.
\par\indent\rlap{\rm(c)}\hphantom{\rm(a)}
$\Prob[v<x_k^*]\le e^{-2g^2\!/s}$ if\/ $i_v=\lceil ks/n+g\rceil$.
\par\indent\rlap{\rm(d)}\hphantom{\rm(a)}
$\Prob[x_{k_r}^*<v]\le e^{-2g^2\!/s}$.
\par\indent\rlap{\rm(e)}\hphantom{\rm(a)}
$i_u\ne\lceil ks/n-g\rceil$ iff\/ $k\le gn/s${\rm;}
$i_v\ne\lceil ks/n+g\rceil$ iff\/ $n<k+gn/s$.
\end{lemma}
\begin{proof}
(a) If $x_k^*<y_{i_u}^*$, at least $s-i_u+1$ samples satisfy
$y_i\ge x_{\bar\jmath+1}^*$ with $\bar\jmath:=\max_{x_j^*=x_k^*}j$.
In the setting of Fact \ref{f:balls}, we have $r:=n-\bar\jmath$ red
elements $x_j\ge x_{\bar\jmath+1}^*$, $ps=s-\bar\jmath s/n$ and
$r'\ge s-i_u+1$.  Since $i_u=\lceil ks/n-g\rceil<ks/n-g+1$ and
$\bar\jmath\ge k$, we get $r'>ps+(\bar\jmath-k)s/n+g\ge ps+g$.
Hence $\Prob[x_k^*<u]\le\Prob[r'\ge ps+g]\le e^{-2g^2\!/s}$
by \eqref{Pexpg}.

(b) If $y_{i_u}^*<x_{k_l}^*$, $i_u$ samples are at most $x_r^*$, where
$r:=\max_{x_j^*<x_{k_l}^*}j$.  Thus we have $r$ red elements
$x_j\le x_r^*$, $ps=rs/n$ and $r'\ge i_u$.  Now,
$1\le r\le k_l-1$ implies $2\le k_l=\lceil k-2gn/s\rceil$ by
\eqref{klkr} and thus $k_l<k-2gn/s+1$, so $-rs/n>-ks/n+2g$.  Hence
$i_u-ps-g\ge ks/n-g-rs/n-g>0$, i.e., $r'>ps+g$; invoke
\eqref{Pexpg} as before.

(c) If $y_{i_v}^*<x_k^*$, $i_v$ samples are at most $x_r^*$, where
$r:=\max_{x_j^*<x_k^*}j$.  Thus we have $r$ red elements
$x_j\le x_r^*$, $ps=rs/n$ and $r'\ge i_v$.  But
$i_v-ps-g\ge ks/n+g-rs/n-g\ge0$ implies $r'\ge ps+g$, so again
\eqref{Pexpg} yields the conclusion.

(d) If $x_{k_r}^*<y_{i_v}^*$, $s-i_v+1$ samples are at least
$x_{\bar\jmath+1}^*$, where $\bar\jmath:=\max_{x_j^*=x_{k_r}^*}j$.
Thus we have $r:=n-\bar\jmath$ red elements
$x_j\ge x_{\bar\jmath+1}^*$, $ps=s-\bar\jmath s/n$ and
$r'\ge s-i_v+1$.  Now, $i_v<ks/n+g+1$ and
$\bar\jmath\ge k_r\ge k+2gn/s$ (cf.\ \eqref{klkr}) yield
$s-i_v+1-ps-g\ge\bar\jmath s/n-ks/n-g-1+1-g\ge0$.  Thus
$x_{k_r}^*<v$ implies $r'\ge ps+g$; hence
$\Prob[x_{k_r}^*<v]\le\Prob[r'\ge ps+g]\le e^{-2g^2\!/s}$
by \eqref{Pexpg}.

(e) Follows immediately from the properties of $\lceil\cdot\rceil$
\cite[\S1.2.4]{knu:acpI3}.
\qed
\end{proof}

We may now estimate the partitioning costs of Step 4.  We assume
that only necessary comparisons are made (but it will be seen that up
to $s$ extraneous comparisons may be accomodated in our analysis;
cf.\ Rem.\ \ref{r:binsample}(a)).
%
%   *** LEMMA 3.4 ***
\begin{lemma}
\label{l:comp}
Let\/ $c$ denote the number of comparisons made at Step\/ $4$.
Then\/
\begin{subequations}
\label{PEcomp}
\begin{equation}
\Prob[\,c\le\bar c\,]\ge1-e^{-2g^2\!/s}\quad\mbox{and}\quad
\Exp c\le\bar c+2(n-s)e^{-2g^2\!/s}\quad\mbox{with}\quad
\label{PEcomp:a}
\end{equation}
\begin{equation}
\bar c:=n+\min\{\,k,n-k\,\}-s+2gn/s.
\label{PEcomp:b}
\end{equation}
\end{subequations}
%In general, $c\le2(n-s)$.
\end{lemma}
\begin{proof}
Consider the event ${\cal A}:=\{c\le\bar c\}$ and its complement
${\cal A}':=\{c>\bar c\}$.  If $u=v$ then $c=n-s\le\bar c$; hence
$\Prob[{\cal A}']=\Prob[{\cal A}'\cap\{u<v\}]$, and we may assume
$u<v$ below.

First, suppose $k<n/2$.  Then
$c=n-s+|\{x\in X\setminus S:x<v\}|$, since $n-s$ elements of
$X\setminus S$ are compared to $v$ first.  In particular, $c\le2(n-s)$.
Since $k<n/2$, $\bar c=n+k-s+2gn/s$.  If
$v\le x_{k_r}^*$, then
$\{x\in X\setminus S:x<v\}\subset\{x\in X:x\le v\}\setminus\{u,v\}$
yields $|\{x\in X\setminus S:x<v\}|\le k_r-2$, so
$c\le n-s+k_r-2$; since $k_r<k+2gn/s+1$, we get
$c\le n+k-s+2gn/s-1\le\bar c$.
Thus $u<v\le x_{k_r}^*$ implies ${\cal A}$.  Therefore,
${\cal A}'\cap\{u<v\}$ implies $\{x_{k_r}^*<v\}\cap\{u<v\}$, so
$\Prob[{\cal A}'\cap\{u<v\}]\le\Prob[x_{k_r}^*<v]\le e^{-2g^2\!/s}$
(Lem.\ \ref{l:rank}(d)).  Hence we have \eqref{PEcomp}, since
$\Exp c\le\bar c+2(n-s)e^{-2g^2\!/s}$
by Fact \ref{f:Ebound} (with $z:=c$, $\zeta:=2(n-s)$).

Next, suppose $k\ge n/2$.  Now
$c=n-s+|\{x\in X\setminus S:u<x\}|$, since $n-s$ elements of
$X\setminus S$ are compared to $u$ first.  If $x_{k_l}^*\le u$, then
$\{x\in X\setminus S:u<x\}\subset\{x\in X:u\le x\}\setminus\{u,v\}$
yields $|\{x\in X\setminus S:u<x\}|\le n-k_l-1$; hence
$k_l\ge k-2gn/s$ gives $c\le n-s+(n-k)+2gn/s-1\le\bar c$.  Thus
${\cal A}'\cap\{u<v\}$ implies $\{u<x_{l_r}^*\}\cap\{u<v\}$, so
$\Prob[{\cal A}'\cap\{u<v\}]\le\Prob[u<x_{k_l}^*]\le e^{-2g^2\!/s}$
(Lem.\ \ref{l:rank}(b)), and we get \eqref{PEcomp} as before.
\qed
\end{proof}

The following result will imply that, for suitable choices of $s$
and $g$, the set $\hat X$ selected at Step 6 will be ``small enough''
with high probability and in expectation; we let $\hat X:=\emptyset$ and
$\hat n:=0$ if Step 5 returns $u$ or $v$, but we don't consider this
case explicitly.
%
%   *** LEMMA 3.5 ***
\begin{lemma}
\label{l:PX}
$\Prob\left[\hat n<4gn/s\right]\ge1-4e^{-2g^2\!/s}$, and\/
$\Exp\hat n\le4gn/s+4ne^{-2g^2\!/s}$.
\end{lemma}
\begin{proof}
The first bound yields the second one by Fact \ref{f:Ebound}
(with $z:=\hat n<n$).  In each case below, we define an event
${\cal E}$ that implies the event ${\cal B}:=\{\hat n<4gn/s\}$.

First, consider the {\em middle\/} case of $i_u=\lceil ks/n-g\rceil$
and $i_v=\lceil ks/n+g\rceil$.  Let
${\cal E}:=\{x_{k_l}^*\le u\le x_k^*\le v\le x_{k_r}^*\}$.  By
Lem.\ \ref{l:rank} and the Boole-Benferroni inequality, its complement
${\cal E}'$ has $\Prob[{\cal E}']\le4e^{-2g^2\!/s}$, so
$\Prob[{\cal E}]\ge1-4e^{-2g^2\!/s}$.  By the rules of Steps 4--6,
$u\le x_k^*\le v$ implies $\hat X=C$, whereas
$x_{k_l}^*\le u\le v\le x_{k_r}^*$ yields
$\hat n\le k_r-k_l+1-2$; since $k_r<k+2gn/s+1$ and $k_l\ge k-2gn/s$
by \eqref{klkr}, we get $\hat n<4gn/s$.  Hence ${\cal E}\subset{\cal B}$
and thus $\Prob[{\cal B}]\ge\Prob[{\cal E}]$.

Next, consider the {\em left\/} case of
$i_u\ne\lceil ks/n-g\rceil$, i.e., $k\le gn/s$ (Lem.\ \ref{l:rank}(e)).
If $i_v\ne\lceil ks/n+g\rceil$, then $n<k+gn/s$
(Lem.\ \ref{l:rank}(e)) gives $\hat n<n<k+gn/s\le2gn/s$; take
${\cal E}:=\{n<k+gn/s\}$, a certain event.  For
$i_v=\lceil ks/n+g\rceil$, let
${\cal E}:=\{x_k^*\le v\le x_{k_r}^*\}$; again
$\Prob[{\cal E}]\ge1-2e^{-2g^2\!/s}$ by Lem.\ \ref{l:rank}(c,d).  Now,
$x_k^*\le v$ implies $\hat X\subset A\cup C$, whereas $v\le x_{k_r}^*$
gives $\hat n\le k_r-1<k+2gn/s\le3gn/s$; therefore
${\cal E}\subset{\cal B}$.

Finally, consider the {\em right\/} case of
$i_v\ne\lceil ks/n+g\rceil$, i.e., $n<k+gn/s$.  If
$i_u\ne\lceil ks/n-g\rceil$ then $k\le gn/s$ gives $\hat n<n<2gn/s$;
take ${\cal E}:=\{k\le gn/s\}$.  For $i_u=\lceil ks/n-g\rceil$,
${\cal E}:=\{x_{k_l}^*\le u\le x_k^*\}$ has
$\Prob[{\cal E}]\ge1-2e^{-2g^2\!/s}$ by Lem.\ \ref{l:rank}(a,b).  Now,
$u\le x_k^*$ implies $\hat X\subset C\cup E$, whereas $x_{k_l}^*\le u$
yields $\hat n\le n-k_l$ with $k_l\ge k-2gn/s$ and thus
$\hat n<3gn/s$.  Hence ${\cal E}\subset{\cal B}$.
\qed
\end{proof}
%
%   *** COROLLARY 3.6 ***
\begin{corollary}
\label{c:PcX}
$\Prob\left[c\le\bar c\ \mbox{and}\ \hat n<4gn/s\right]\ge
1-4e^{-2g^2\!/s}$.
\end{corollary}
\begin{proof}
Check that ${\cal E}$ implies ${\cal A}$ in the proofs of
Lems.\ \ref{l:comp} and \ref{l:PX}; note that
$n\le2gn/s$ yields $c\le2(n-s)\le\bar c$ (cf.\ \eqref{PEcomp:b}) in
the left and right subcases.
\qed
\end{proof}
%
%   *** REMARK 3.7 ***
\begin{remark}
\label{r:PcX}
\rm
Suppose Step 3 resets $i_u:=i_v$ if $k\le gn/s$, or $i_v:=i_u$ if
$n<k+gn/s$, finding a single pivot $u=v$ in these cases.  The preceding
results remain valid. % for this modification.
\end{remark}
%
%   *** SECTION 4 ***
\section{Analysis of the recursive version}
\label{s:analrec}
In this section we analyze the average performance of {\sc Select} for
various sample sizes.
%
%   *** SUBSECTION 4.1 ***
\subsection{Floyd-Rivest's samples}
\label{ss:FRsample}
For positive constants $\alpha$ and $\beta$, consider choosing
$s=s(n)$ and $g=g(n)$ as
\begin{equation}
s:=\min\left\{\lceil\alpha f(n)\rceil,n-1\right\}\ \mbox{and}\
g:=(\beta s\ln n)^{1/2}\ \mbox{with}\ f(n):=n^{2/3}\ln^{1/3}n.
\label{sgf}
\end{equation}
This form of $g$ gives a probability bound
$e^{-2g^2\!/s}=n^{-2\beta}$ for Lems.\ \ref{l:comp}--\ref{l:PX}.
To get more feeling, suppose $\alpha=\beta=1$ and $s=f(n)$.
Let $\phi(n):=f(n)/n$.  Then $s/n=g/s=\phi(n)$ and $\hat n/n$ is
at most $4\phi(n)$ with high probability (at least $1-4/n^2$), i.e.,
$\phi(n)$ is a contraction factor; note that $\phi(n)\approx2.4\%$ for
$n=10^6$ (cf.\ Tab.\ \ref{tab:fnphin}).
%
%   *** TABLE 4.1 ***
\begin{table}
\caption{Sample size $f(n):=n^{2/3}\ln^{1/3}n$ and relative sample size
$\phi(n):=f(n)/n$.}
\label{tab:fnphin}
\footnotesize
\begin{center}
\begin{tabular}{ccccccccc}
\hline
\vphantom{$1^{2^3}$} % Need more vertical space!
$n$     & $10^3$ & $10^4$ & $10^5$ & $10^6$ & $5\cdot10^6$ & $10^7$
        & $5\cdot10^7$    & $10^8$ \\
\hline
$f(n)$  & 190.449& 972.953& 4864.76& 23995.0&       72287.1& 117248
        & 353885 & 568986 \\
$\phi(n)$
        & .190449& .097295& .048648& .023995&       .014557& .011725
        & .007078& .005690\\
\hline
\end{tabular}
\end{center}
\end{table}
%
%   *** THEOREM 4.1 ***
\begin{theorem}
\label{t:selFR}
Let\/ $C_{nk}$ denote the expected number of comparisons made by
{\sc Select} for $s$ and\/ $g$ chosen as in\/ \eqref{sgf} with\/
$\beta\ge1/6$.  There exists a positive constant\/ $\gamma$ such
that
\begin{equation}
C_{nk}\le n+\min\{\,k,n-k\,\}+\gamma f(n)\quad\forall1\le k\le n.
\label{CnkFR}
\end{equation}
\end{theorem}
\begin{proof}
We need a few preliminary facts.
The function $\phi(t):=f(t)/t=(\ln t/t)^{1/3}$ decreases to $0$ on
$[e,\infty)$, whereas $f(t)$ grows to infinity on $[2,\infty)$.
Let $\delta:=4(\beta/\alpha)^{1/2}$.
Pick $\bar n\ge3$ large enough so that
$e-1\le\alpha f(\bar n)\le\bar n-1$ and $e\le\delta f(\bar n)$.
Let $\bar\alpha:=\alpha+1/f(\bar n)$.
Then, by \eqref{sgf} and the monotonicity of $f$ and $\phi$, we have
for $n\ge\bar n$
\begin{equation}
s\le\bar\alpha f(n)\quad\mbox{and}\quad
f(s)\le\bar\alpha\phi(\bar\alpha f(\bar n))f(n),
\label{sfsFR}
\end{equation}
\begin{equation}
f(\delta f(n))\le\delta\phi(\delta f(\bar n))f(n).
\label{fdeltaFR}
\end{equation}
For instance, the first inequality of \eqref{sfsFR} yields
$f(s)\le f(\bar\alpha f(n))$, whereas
$$
f(\bar\alpha f(n))=\bar\alpha\phi(\bar\alpha f(n))f(n)\le
\bar\alpha\phi(\bar\alpha f(\bar n))f(n).
$$
Also for $n\ge\bar n$,
we have $s=\lceil\alpha f(n)\rceil=\alpha f(n)+\epsilon$ with
$\epsilon\in[0,1)$ in \eqref{sgf}.  Writing $s=\tilde\alpha f(n)$ with
$\tilde\alpha:=\alpha+\epsilon/f(n)\in[\alpha,\bar\alpha)$, we deduce
from \eqref{sgf} that
\begin{equation}
gn/s=(\beta/\tilde\alpha)^{1/2}f(n)\le(\beta/\alpha)^{1/2}f(n).
\label{gnsboundFR}
\end{equation}
In particular, $4gn/s\le\delta f(n)$, since
$\delta:=4(\beta/\alpha)^{1/2}$.
For $\beta\ge1/6$, \eqref{sgf} implies
\begin{equation}
ne^{-2g^2\!/s}\le
n^{1-2\beta}=f(n)n^{1/3-2\beta}\ln^{-1/3}n\le f(n)\ln^{-1/3}n.
\label{ne2g2sFR}
\end{equation}
Using the monotonicity of $\phi$ and $f$ on $[e,\infty)$, increase
$\bar n$ if necessary to get
\begin{equation}
2\bar\alpha\phi(\bar\alpha f(\bar n))+\delta\phi(\delta f(\bar n))+
4\phi(\bar n)\bar n^{1/3-2\beta}\ln^{-1/3}\bar n\le0.95.
\label{kapreqFR}
\end{equation}
By Rem.\ \ref{r:selABCDE}(c), there is $\gamma$ such that \eqref{CnkFR}
holds for all $n\le\bar n$; increasing $\gamma$ if necessary, we have
\begin{equation}
2\bar\alpha+2\delta+8\bar n^{1/3-2\beta}\ln^{-1/3}\bar n\le0.05\gamma.
\label{gamreqFR}
\end{equation}

Let $n'\ge\bar n$.  Assuming \eqref{CnkFR} holds for all $n\le n'$,
for induction let $n=n'+1$.

The cost of Step 3 can be estimated as follows.
We may first apply {\sc Select} recursively to $S$ to find
$u=y_{i_u}^*$, and then extract $v=y_{i_v}^*$ from the elements
$y_{i_u+1}^*,\ldots,y_s^*$ (assuming $i_u<i_v$; otherwise $v=u$).
Since $s\le n'$, the expected number of comparisons is
\begin{equation}
C_{si_u}+C_{s-i_u,i_v-i_u}\le
1.5s+\gamma f(s)+1.5(s-i_u)+\gamma f(s-i_u)
\le3s-1.5+2\gamma f(s).
\label{CsiuCFR}
\end{equation}

The partitioning cost of Step 4 is estimated by \eqref{PEcomp} as
\begin{equation}
\Exp c\le n+\min\{\,k,n-k\,\}-s+2gn/s+2ne^{-2g^2\!/s}.
\label{EcompFR}
\end{equation}

The cost of finishing up at Step 7 is at most
$C_{\hat n\hat k}\le1.5\hat n+\gamma f(\hat n)$.  But by
Lem.\ \ref{l:PX}, $\Prob[\hat n\ge4gn/s]\le4e^{-2g^2\!/s}$,
and $\hat n<n$, so
(cf.\ Fact \ref{f:Ebound} with $z:=1.5\hat n+\gamma f(\hat n)$)
$$
\Exp\left[\,1.5\hat n+\gamma f(\hat n)\,\right]\le
1.5\cdot4gn/s+\gamma f(4gn/s)+
\left[\,1.5n+\gamma f(n)\,\right]4e^{-2g^2\!/s}.
$$
Since $4gn/s\le\delta f(n)$, $f$ is increasing, and $f(n)=\phi(n)n$
above, we get
\begin{equation}
\Exp C_{\hat n\hat k}\le
6gn/s+\gamma f(\delta f(n))+
\left[\,1.5+\gamma\phi(n)\,\right]4ne^{-2g^2\!/s}.
\label{ECFR}
\end{equation}

Add the costs \eqref{CsiuCFR}, \eqref{EcompFR} and \eqref{ECFR} to get
\begin{subequations}
\label{Cnk2sFR}
\begin{eqnarray}
C_{nk}&\le&3s-1.5+2\gamma f(s)+
n+\min\{\,k,n-k\,\}-s+2gn/s+2ne^{-2g^2\!/s}\nonumber\\
&&\quad{}+6gn/s+\gamma f(\delta f(n))+
\left[\,1.5+\gamma\phi(n)\,\right]4ne^{-2g^2\!/s}\nonumber\\
\label{Cnk2sFR:a}
&\le&n+\min\{\,k,n-k\,\}+
\left[\,2s+8gn/s+8ne^{-2g^2\!/s}\,\right]\\
\label{Cnk2sFR:b}
&&\quad{}+\gamma\left[\,2f(s)+f(\delta f(n))+
4ne^{-2g^2\!/s}\phi(n)\,\right].
\end{eqnarray}
\end{subequations}
By \eqref{sfsFR}--\eqref{ne2g2sFR}, the bracketed term in
\eqref{Cnk2sFR:a} is at most $0.05\gamma f(n)$ due to
\eqref{gamreqFR}, and that in \eqref{Cnk2sFR:b} is at most $0.95f(n)$
from \eqref{kapreqFR}; thus \eqref{CnkFR} holds as required.
\qed
\end{proof}

We now indicate briefly how to adapt the preceding proof to several
variations on \eqref{sgf}; choices similar to \eqref{sgfFRlns} and
\eqref{sgfFRsn2/3} are used in \cite{meh:osa} and \cite{flri:asf},
respectively.
%
%   *** REMARKS 4.2 ***
\begin{remarks}
\label{r:selFR}
\rm
(a)
Theorem \ref{t:selFR} holds for the following modification of \eqref{sgf}:
\begin{equation}
s:=\min\left\{\lceil\alpha f(n)\rceil,n-1\right\}\ \mbox{and}\
g:=(\beta s\ln\theta s)^{1/2}\ \mbox{with}\ f(n):=n^{2/3}\ln^{1/3}n,
\label{sgfFRlns}
\end{equation}
provided that $\beta\ge1/4$, where $\theta>0$.  Indeed, the analogue of
\eqref{gnsboundFR} (cf.\ \eqref{sgf}, \eqref{sgfFRlns})
\begin{equation}
gn/s=(\beta/\tilde\alpha)^{1/2}f(n)(\ln\theta s/\ln n)^{1/2}\le
(\beta/\alpha)^{1/2}f(n)(\ln\theta s/\ln n)^{1/2}
\label{gnsboundFRlns}
\end{equation}
works like \eqref{gnsboundFR} for large $n$ (since
$\lim_{n\to\infty}\frac{\ln\theta s}{\ln n}=2/3$),
whereas replacing \eqref{ne2g2sFR} by
\begin{equation}
ne^{-2g^2\!/s}=n(\theta s)^{-2\beta}\le
f(n)(\alpha\theta)^{-2\beta}n^{(1-4\beta)/3}\ln^{-(1+2\beta)/3}n,
\label{ne2g2sFRlns}
\end{equation}
we may replace $\bar n^{1/3-2\beta}$ by
$(\alpha\theta)^{-2\beta}\bar n^{(1-4\beta)/3}$
in \eqref{kapreqFR}--\eqref{gamreqFR}.
\par(b)
Theorem \ref{t:selFR} holds for the following modification of \eqref{sgf}:
\begin{equation}
s:=\min\left\{\lceil\alpha f(n)\rceil,n-1\right\}\ \mbox{and}\
g:=(\beta s\ln^{\epsilon_l}n)^{1/2}\ \mbox{with}\
f(n):=n^{2/3}\ln^{\epsilon_l/3}n,
\label{sgfFRlneps}
\end{equation}
provided either $\epsilon_l=1$ and $\beta\ge1/6$, or $\epsilon_l>1$.
Indeed, since \eqref{sgfFRlneps}$=$\eqref{sgf} for $\epsilon_l=1$,
suppose $\epsilon_l>1$.  Clearly, \eqref{sfsFR}--\eqref{gnsboundFR}
hold with $\phi(t):=f(t)/t$.  For $\tilde\beta\ge1/6$ and $n$ large
enough, we have $g^2\!/s=\beta\ln^{\epsilon_l}n\ge\tilde\beta\ln n$;
hence, replacing $2\beta$ by $2\tilde\beta$ and $\ln^{-1/3}$ by
$\ln^{-\epsilon_l/3}$ in \eqref{ne2g2sFR}--\eqref{gamreqFR}, we may
use the proof of Thm \ref{t:selFR}.
\par(c)
Theorem \ref{t:selFR} remains true if we use $\beta\ge1/6$,
\begin{equation}
s:=\min\left\{\left\lceil\alpha n^{2/3}\right\rceil,n-1\right\},\
g:=(\beta s\ln n)^{1/2}\ \mbox{and}\
f(n):=n^{2/3}\ln^{1/2}n.
\label{sgfFRsn2/3}
\end{equation}
Again \eqref{sfsFR}--\eqref{gnsboundFR} hold with $\phi(t):=f(t)/t$,
and $\ln^{-1/2}$ replaces $\ln^{-1/3}$ in
\eqref{ne2g2sFR}--\eqref{gamreqFR}.
\par(d)
None of these choices gives $f(n)$ better than that in \eqref{sgf} for
the bound \eqref{CnkFR}.
\end{remarks}
%
%   *** SUBSECTION 4.2 ***
\subsection{Reischuk's samples}
\label{ss:Rsample}
For positive constants $\alpha$ and $\beta$, consider using
\begin{subequations}
\label{sgR}
\begin{equation}
s:=\min\left\{\,\lceil\alpha n^{\epsilon_s}\rceil,n-1\,\right\}
\quad\mbox{and}\quad
g:=\left(\,\beta sn^{\epsilon}\,\right)^{1/2}\quad\mbox{with}\quad
\label{sgR:a}
\end{equation}
\begin{equation}
\eta:=\max\left\{\,1+(\epsilon-\epsilon_s)/2,\epsilon_s\,\right\}<1
\quad\mbox{for some fixed}\ 0<\epsilon<\epsilon_s.
\label{sgR:b}
\end{equation}
\end{subequations}
%
%   *** THEOREM 4.3 ***
\begin{theorem}
\label{t:selR}
Let\/ $C_{nk}$ denote the expected number of comparisons made
by {\sc Select} for\/ $s$ and\/ $g$ chosen as in\/ \eqref{sgR}.
There exists a positive constant\/ $\gamma_\eta$ such that for
all\/ $k\le n$
\begin{equation}
C_{nk}\le n+\min\{\,k,n-k\,\}+\gamma_\eta f_\eta(n)
\quad\mbox{with}\quad f_\eta(n):=n^{\eta}.
\label{CnkR}
\end{equation}
\end{theorem}
\begin{proof}
The function $f_\eta(t):=t^{\eta}$ grows to $\infty$ on
$(0,\infty)$, whereas $\phi_\eta(t):=f_\eta(t)/t=t^{\eta-1}$ decreases
to $0$, so $f_\eta$ and $\phi_\eta$ may replace $f$ and $\phi$ in the
proof of Thm \ref{t:selFR}.  Indeed, picking $\bar n\ge1$ such that
$\alpha\bar n^{\epsilon_s}\le\bar n-1$, for $n\ge\bar n$ we may use
$s=\tilde\alpha n^{\epsilon_s}\le\bar\alpha f_\eta(n)$ with
$\alpha\le\tilde\alpha\le\bar\alpha:=1+1/\bar n^{\epsilon_s}$ to get
analogues \eqref{sfsFR}--\eqref{fdeltaFR} and the following analogue of
\eqref{gnsboundFR}
\begin{equation}
gn/s=(\beta/\tilde\alpha)^{1/2}n^{1+(\epsilon-\epsilon_s)/2}\le
(\beta/\alpha)^{1/2}f_\eta(n).
\label{gnsboundR}
\end{equation}
Since $g^2\!/s=\beta n^{\epsilon}$ by \eqref{sgR}, and
$te^{-2\beta t^{\epsilon}}\!\!/t^{\eta}$ decreases to
$0$ for $t\ge t_\eta:=
\left(\frac{1-\eta}{2\beta\epsilon}\right)^{1/\epsilon}$,
we may replace \eqref{ne2g2sFR} by
\begin{equation}
ne^{-2g^2\!/s}=ne^{-2\beta n^{\epsilon}}\le
\bar n^{1-\eta}e^{-2\beta\bar n^{\epsilon}}
f_\eta(n)\quad\forall n\ge\bar n\ge t_\eta.
\label{ne2g2sR}
\end{equation}
Hence, with $\bar n^{1-\eta}e^{-2\beta\bar n^{\epsilon}}$
replacing $\bar n^{1/3-2\beta}\ln^{-1/3}\bar n$ in
\eqref{kapreqFR}--\eqref{gamreqFR}, the proof goes through.
\qed
\end{proof}
%
%   *** REMARKS 4.4 ***
\begin{remarks}
\label{r:selR}
\rm
(a)
For a fixed $\epsilon\in(0,1)$, minimizing $\eta$ in \eqref{sgR}
yields the {\em optimal\/} sample size parameter
\begin{equation}
\epsilon_s:=(2+\epsilon)/3,
\label{epssRopt}
\end{equation}
with $\eta=\epsilon_s>2/3$ and $f_\eta(n)=n^{(2+\epsilon)/3}$; note
that if $s=\alpha n^{\epsilon_s}$ in \eqref{sgR}, then
$g=(\alpha\beta)^{1/2}n^{\epsilon_g}$ with
$\epsilon_g:=(1+2\epsilon)/3$.
To compare the bounds \eqref{CnkFR} and \eqref{CnkR} for this optimal
choice, let $\Phi_\epsilon(t):=(t^\epsilon\!/\ln t)^{1/3}$, so that
$\Phi_\epsilon(t)=f_\eta(t)/f(t)=\phi_\eta(t)/\phi(t)$.  Since
$\lim_{n\to\infty}\Phi_\epsilon(n)=\infty$, the choice \eqref{sgf}
is asymptotically superior to \eqref{sgR}.  However,
$\Phi_\epsilon(n)$ grows quite slowly, and $\Phi_\epsilon(n)<1$ even
for fairly large $n$ when $\epsilon$ is small
(cf.\ Tab.\ \ref{tab:FR/R}).
%
%   *** TABLE 4.2 ***
\begin{table}
\caption{Relative sample sizes $\Phi_{\epsilon}(n)$ and probability
bounds $e^{-2n^{\epsilon}}$.}
\label{tab:FR/R}
\footnotesize
\begin{center}
\begin{tabular}{rr|cccc|llll}
\hline
  & &\multicolumn{4}{c|}{$\Phi_{\epsilon}(n):=(t^\epsilon\!/\ln t)^{1/3}$%
     \vphantom{$1^{2^3}$}} % Need more vertical space!
    &\multicolumn{4}{c}{$\exp(-2n^{\epsilon})$}\\
%\hline
$n$ &   & $10^5$  & $10^6$  & $5\cdot10^6$ & $10^7$%
\vphantom{$1^{2^3}$} % Need more vertical space!
    &\multicolumn{1}{|c}{$10^5$}
    &\multicolumn{1}{c}{$10^6$}
    &\multicolumn{1}{c}{$5\cdot10^6$}
    &\multicolumn{1}{c}{$10^7$}\\
\hline
     \vphantom{$1^{2^3}$} % Need more vertical space!
 &$1/4$ & 1.16               & 1.32               & 1.45      & 1.52
        & $3.6\cdot10^{-16}$ & $3.4\cdot10^{-28}$ & $8.4\cdot10^{-42}$
        & $1.4\cdot10^{-49}$\\
$\epsilon$
 & $1/6$& .840               & .898               & .946      & .969
        & $1.2\cdot10^{-6}$  & $2.1\cdot10^{-9}$  & $4.4\cdot10^{-12}$
        & $1.8\cdot10^{-12}$\\
 &$1/9$ & .678               & .695               & .711      & .719
        & $7.6\cdot10^{-4}$  & $9.3\cdot10^{-5}$  & $1.5\cdot10^{-5}$
        & $6.2\cdot10^{-6}$\\
\hline
\end{tabular}
\end{center}
\end{table}
On the other hand, for small $\epsilon$ and $\beta=1$, the probability
bound $e^{-2g^2\!/s}=e^{-2n^\epsilon}$ of \eqref{sgR} is weak relative
to $e^{-2g^2\!/s}=n^{-2}$ ensured by \eqref{sgf}.
\par(b)
Consider using
$s:=\min\{\lceil\alpha n^{\epsilon_s}\rceil,n-1\}$ and
$g:=\beta^{1/2}n^{\epsilon_g}$ with $\epsilon_s,\epsilon_g\in(0,1)$
such that $\epsilon:=2\epsilon_g-\epsilon_s>0$ and
$\eta:=\max\{1+\epsilon_g-\epsilon_s,\epsilon_s\}<1$.
Theorem \ref{t:selR} covers this choice.  Indeed, the equality
$1+\epsilon_g-\epsilon_s=1+(\epsilon-\epsilon_s)/2$ shows that
\eqref{sgR:b} and \eqref{gnsboundR} remain valid, and we have the
following analogue of \eqref{ne2g2sR}
\begin{equation}
ne^{-2g^2\!/s}\le
\bar n^{1-\eta}e^{-2(\beta/\bar\alpha)\bar n^{\epsilon}}
f_\eta(n)\quad\forall n\ge\bar n\ge % t_\eta:=
[(1-\eta)\bar\alpha/(2\beta\epsilon)]^{1/\epsilon},
\label{ne2g2sRgen}
\end{equation}
so compatible modifications of \eqref{kapreqFR}--\eqref{gamreqFR}
suffice for the rest of the proof.  Note that
$\eta\ge(2+\epsilon)/3$ by (a); for the choice $\epsilon_s=\frac12$,
$\epsilon_g=\frac7{16}$ of \cite{rei:ppa},
$\epsilon=\frac38$ and $\eta=\frac{15}{16}$.
\end{remarks}
%
%   *** SUBSECTION 4.3 ***
\subsection{Handling small subfiles}
\label{ss:subfile}
Since the sampling efficiency decreases when $X$ shrinks, consider the
following modification.  For a fixed cut-off parameter
$n_{\rm cut}\ge1$, let sSelect$(X,k)$ be a ``small-select'' routine that
finds the $k$th smallest element of $X$ in at most $C_{\rm cut}<\infty$
comparisons when $|X|\le n_{\rm cut}$ (even bubble sort will do).  Then
{\sc Select} is modified to start with the following
\medbreak\noindent{\bf Step 0} ({\em Small file case\/}).
If $n:=|X|\le n_{\rm cut}$, return sSelect$(X,k)$.

Our preceding results remain valid for this modification.  In fact it
suffices if $C_{\rm cut}$ bounds the {\em expected\/} number of
comparisons of sSelect$(X,k)$ for $n\le n_{\rm cut}$.  For instance,
\eqref{CnkFR} holds for $n\le n_{\rm cut}$ and $\gamma\ge C_{\rm cut}$,
and by induction as in Rem.\ \ref{r:selABCDE}(c) we have $C_{nk}<\infty$
for all $n$, which suffices for the proof of Thm \ref{t:selFR}.

Another advantage is that even small $n_{\rm cut}$ ($1000$ say) limits
nicely the stack space for recursion.  Specifically, the tail
recursion of Step 7 is easily eliminated (set $X:=\hat X$, $k:=\hat k$
and go to Step 0), and the calls of Step 3 deal with subsets whose
sizes quickly reach $n_{\rm cut}$.  For example, for the choice of
\eqref{sgf} with $\alpha=1$ and $n_{\rm cut}=600$, at most four
recursive levels occur for $n\le2^{31}\approx2.15\cdot10^9$.
%
%   *** SECTION 5 ***
\section{Analysis of nonrecursive versions}
\label{s:analnonrec}
Consider a nonrecursive version of {\sc Select} in which Steps 3 and 7,
instead of {\sc Select}, employ a linear-time routine (e.g., {\sc Pick}
\cite{blflprrita:tbs}) that finds the $i$th smallest of $m$ elements
in at most $\gamma_Pm$ comparisons for some constant $\gamma_P>2$.
%
%   *** THEOREM 5.1 ***
\begin{theorem}
\label{t:selnonrec}
Let\/ $c_{nk}$ denote the number of comparisons made by the nonrecursive
version of\/ {\sc Select} for a given choice of\/ $s$ and\/ $g$.
Suppose\/ $s<n-1$.
\par\indent\rlap{\rm(a)}\hphantom{\rm(a)}
For the choice of\/ \eqref{sgf} with\/ $f(n):=n^{2/3}\ln^{1/3}n$, we have
\begin{subequations}
\label{cknFR}
\begin{equation}
\Prob\left[\,c_{nk}\le n+\min\{\,k,n-k\,\}+\hat\gamma_Pf(n)\,\right]\ge
1-4n^{-2\beta}\quad\mbox{with}
\label{cknFR:a}
\end{equation}
\begin{equation}
\hat\gamma_P:=(4\gamma_P+2)(\beta/\alpha)^{1/2}+
(2\gamma_P-1)\left[\alpha+1/f(n)\right],
\label{cknFR:b}
\end{equation}
\end{subequations}
also with\/ $f(n)$ in\/ \eqref{cknFR:b} replaced by\/ $f(3)>2$
{\rm(}since\/ $n\ge3${\rm)}.
Moreover, if\/ $\beta\ge1/6$, then
\begin{equation}
\Exp c_{nk}\le n+\min\{\,k,n-k\,\}+
\left(\,\hat\gamma_P+4\gamma_P+2\,\right)f(n).
\label{EcknFR}
\end{equation}
\par\indent\rlap{\rm(b)}\hphantom{\rm(a)}
For the choice of\/ \eqref{sgfFRlns}, if\/ $\theta s\le n$, then\/
\eqref{cknFR:a} holds with\/ $n^{-2\beta}$ replaced by\/
$(\alpha\theta)^{-2\beta}n^{-4\beta/3}\ln^{-2\beta/3}n$.
Moreover, if\/ $\beta\ge1/4$, then\/ \eqref{EcknFR} holds with\/
$4\gamma_P+2$ replaced by\/ $(4\gamma_P+2)(\alpha\theta)^{-2\beta}$.
\par\indent\rlap{\rm(c)}\hphantom{\rm(a)}
For the choice of\/ \eqref{sgR}, \eqref{cknFR} holds with\/
$f(n)$ replaced by\/ $f_\eta(n):=n^\eta$ and\/
$n^{-2\beta}$ by\/ $e^{-2\beta n^{\epsilon}}$.
Moreover, if\/ %$n$ is large enough so that\/
$n^{1-\eta}e^{-2\beta n^{\epsilon}}\le1$, then\/ \eqref{EcknFR}
holds with\/ $f$ replaced by\/ $f_\eta$.
\end{theorem}
\begin{proof}
The cost $c_{nk}$ of Steps 3, 4 and 7 is at most
$2\gamma_Ps+c+\gamma_P\hat n$.  By Cor.\ \ref{c:PcX}, the event
${\cal C}:=\{c\le\bar c,\hat n<4gn/s\}$ has probability
$\Prob[{\cal C}]\ge1-4e^{-2g^2\!/s}$.  If ${\cal C}$ occurs, then
\begin{eqnarray}
c_{nk}&\le&n+\min\{\,k,n-k\,\}-s+2gn/s+2\gamma_Ps+
\gamma_P\lfloor4gn/s\rfloor\nonumber\\
&\le&n+\min\{\,k,n-k\,\}+\left(\,4\gamma_P+2\,\right)gn/s+
\left(\,2\gamma_P-1\,\right)s.
\label{cknbound}
\end{eqnarray}
Similarly, since $\Exp c_{nk}\le2\gamma_Ps+\Exp c+\gamma_P\Exp\hat n$,
Lems.\ \ref{l:comp}--\ref{l:PX} yield
\begin{equation}
\Exp c_{nk}\le n+\min\{\,k,n-k\,\}+\left(\,4\gamma_P+2\,\right)gn/s+
\left(\,2\gamma_P-1\,\right)s+
\left(\,4\gamma_P+2\,\right)ne^{-2g^2\!/s}.
\label{Ecknbound}
\end{equation}

(a) Since $e^{-2g^2\!/s}=n^{-2\beta}$,
$s=\lceil\alpha f(n)\rceil\le\bar\alpha f(n)$ from $s<n-1$ and
\eqref{sfsFR}, and $gn/s$ is bounded by \eqref{gnsboundFR},
\eqref{cknbound} implies \eqref{cknFR}.  Then \eqref{EcknFR} follows
from \eqref{ne2g2sFR} and \eqref{Ecknbound}.

(b) Proceed as for (a), invoking
\eqref{gnsboundFRlns}--\eqref{ne2g2sFRlns}
instead of \eqref{gnsboundFR} and \eqref{ne2g2sFR}.

(c) Argue as for (a), using the proof of Thm \ref{t:selR}, in particular
\eqref{gnsboundR}--\eqref{ne2g2sR}.
\qed
\end{proof}
%
%   *** COROLLARY 5.2 ***
\begin{corollary}
\label{c:selnonrec}
The nonrecursive version of\/ {\sc Select} requires\/
$n+\min\{k,n-k\}+o(n)$ comparisons with probability at least
$1-4n^{-2\beta}$ for the choice of\/ \eqref{sgf}, at least\/
$1-4(\alpha\theta)^{-2\beta}n^{-4\beta/3}$ for the choice of\/
\eqref{sgfFRlns}, and at least $1-4e^{-2\beta n^\epsilon}$ for the
choice of\/ \eqref{sgR}.
\end{corollary}
%
%   *** REMARKS 5.3 ***
\begin{remarks}
\label{r:selnonrec}
\rm
(a)
Suppose Steps 3 and 7 simply sort $S$ and $\hat X$ by any algorithm that
takes at most $\gamma_S(s\ln s+\hat n\ln\hat n)$ comparisons for a
constant $\gamma_S$.  This cost is at most $(s+\hat n)\gamma_S\ln n$,
because $s,\hat n<n$, so we may replace $2\gamma_P$ by
$\gamma_S\ln n$ and $4\gamma_P$ by $4\gamma_S\ln n$ in
\eqref{cknbound}--\eqref{Ecknbound}, and hence in
\eqref{cknFR}--\eqref{EcknFR}.  For the choice of \eqref{sgf}, this
yields
\begin{subequations}
\label{cknsortFR}
\begin{equation}
\Prob\left[\,c_{nk}\le n+\min\{\,k,n-k\,\}+
\hat\gamma_Sf(n)\ln n\,\right]\ge1-4n^{-2\beta}\quad\mbox{with}
\label{cknsortFR:a}
\end{equation}
\begin{equation}
\hat\gamma_S:=(4\gamma_S+2\ln^{-1}n)(\beta/\alpha)^{1/2}+
(\gamma_S-\ln^{-1}n)\left[\alpha+1/f(n)\right],
\label{cknsortFR:b}
\end{equation}
\end{subequations}
\begin{equation}
\Exp c_{nk}\le n+\min\{\,k,n-k\,\}+
\left(\,\hat\gamma_S+4\gamma_S+2\ln^{-1}n\,\right)f(n)\ln n,
\label{EcknsortFR}
\end{equation}
where $\ln^{-1}n$ may be replaced by $\ln^{-1}3$, and \eqref{EcknsortFR}
still needs $\beta\ge1/6$; for the choices \eqref{sgfFRlns} and
\eqref{sgR}, we may modify \eqref{cknsortFR}--\eqref{EcknsortFR} as in
Thm \ref{t:selnonrec}(b,c).
%Then \eqref{EcknFR} yields
%$\Exp c_{nk}\le n+\min\{k,n-k\}+O(f(n)\ln n)$ for the choices
%\eqref{sgf} with $\beta\ge1/6$ and \eqref{sgfFRlns} with $\beta>1/4$,
%with $f(n)$ replaced by $f_\eta(n)$ for the choice \eqref{sgR}.
Corollary \ref{c:selnonrec} remains valid.
\par(b)
The bound \eqref{EcknFR} holds if Steps 3 and 7 employ a routine (e.g.,
{\sc Find} \cite{hoa:fa65}, \cite[\S3.7]{ahhoul:dac})
for which the expected number of comparisons to find the $i$th smallest
of $m$ elements is at most $\gamma_Pm$
(then $\Exp c_{nk}\le2\gamma_Ps+\Exp c+\gamma_P\Exp\hat n$ is bounded
as before).
\par(c)
Suppose Step 6 returns to Step 1 if $\hat n\ge4gn/s$.
By Cor.\ \ref{c:PcX}, such loops are finite wp $1$, and don't
occur with high probability, for $n$ large enough.
\par(d)
Our results improve upon \cite[Thm 2]{gesi:chb}, which only gives an
estimate like \eqref{cknFR:a}, but with $4n^{-2\beta}$ replaced by
$O(n^{1-2\beta/3})$, a much weaker bound.  Further, the approach of
\cite[\S3]{gesi:chb} is restricted to distinct elements.
\end{remarks}

We now comment briefly on the possible use of sampling with
replacement.
%
%   *** REMARKS 5.4 ***
\begin{remarks}
\label{r:binsample}
\rm
(a)
Suppose Step 2 of {\sc Select} employs sampling with replacement.
Since the tail bound \eqref{Pexpg} remais valid for the binomial
distribution \cite{chv:thd,hoe:pis}, Lemma \ref{l:rank} is not
affected.  However, when Step 4 no longer skips comparisons with
the elements of $S$, $-s$ in \eqref{PEcomp} and \eqref{EcompFR} is
replaced by $0$ (cf.\ the proof of Lem.\ \ref{l:comp}), $2s$ in
\eqref{Cnk2sFR:a} by $3s$ and $2\bar\alpha$ in \eqref{gamreqFR}
by $3\bar\alpha$.  Similarly, adding $s$ to the right sides of
\eqref{cknbound}--\eqref{Ecknbound} boils down to omitting $-1$
in \eqref{cknFR:b} and $-\ln^{-1}n$ in \eqref{cknsortFR:b}.  Hence the
preceding results remain valid.
\par(b)
Of course, sampling with replacement needs additional storage for
$S$.  This is inconvenient for the recursive version, but tolerable for
the nonrecursive ones because the sample sizes are relatively small
(hence \eqref{PEcomp} with $-s$ omitted is not too bad).
\par(c)
Our results improve upon \cite[Thm 3.5]{mora:ra}, corresponding to
\eqref{sgR} with $\epsilon=1/4$ and $\beta=1$, where the probability
bound $1-O(n^{-1/4})$ is weaker than our $1-4e^{-2n^{1/4}}$,
sampling is done with replacement and the elements are distinct.
\par(d)
Our results subsume \cite[Thm 2]{meh:osa}, which gives an estimate
like \eqref{EcknFR} for the choice \eqref{sgfFRlns} with $\beta=1$,
using quickselect (cf.\ Rem.\ \ref{r:selnonrec}(b)) and sampling
with replacement in the case of distinct elements.
\end{remarks}
%
%   *** SECTION 6 ***
\section{Ternary and quintary partitioning}
\label{s:ternquint}
In this section we discuss ways of implementing {\sc Select} when
the input set is given as an array $x[1\colon n]$.  We introduce
a modification of the ternary partitioning scheme
\cite{bemc:esf,bese:fas} that obviates subscript range checking,
and give extensions to quintary partitioning.

The following notation is needed to describe the operations of
{\sc Select} in more detail.

Each stage works with a segment $x[l\colon r]$ of the input array
$x[1\colon n]$, where $1\le l\le r\le n$ are such that $x_i<x_l$ for
$i=1\colon l-1$, $x_r<x_i$ for $i=r+1\colon n$, and the $k$th smallest
element of $x[1\colon n]$ is the $(k-l+1)$th smallest element of
$x[l\colon r]$.  The task of {\sc Select} is {\em extended\/}: given
$x[l\colon r]$ and $l\le k\le r$,
{\sc Select}$(x,l,r,k,k_-,k_+)$ permutes $x[l\colon r]$ and finds
$l\le k_-\le k\le k_+\le r$
such that $x_i<x_k$ for all $l\le i<k_-$, $x_i=x_k$ for all
$k_-\le i\le k_+$, $x_i>x_k$ for all $k_+<i\le r$.  The initial call
is {\sc Select}$(x,1,n,k,k_-,k_+)$.

A vector swap denoted by $x[a\colon b]\leftrightarrow x[b+1\colon c]$
means that the first $d:=\min(b+1-a,c-b)$ elements of array
$x[a\colon c]$ are exchanged with its last $d$ elements in arbitrary
order if $d>0$; e.g., we may exchange
$x_{a+i}\leftrightarrow x_{c-i}$ for $0\le i<d$, or
$x_{a+i}\leftrightarrow x_{c-d+1+i}$ for $0\le i<d$.
%
%   *** SUBSECTION 6.1 ***
\subsection{Ternary partitions}
\label{ss:tern}
For a given pivot $v:=x_k$ from the array $x[l\colon r]$, the
following {\em ternary\/} scheme partitions the array into three blocks,
with $x_m<v$ for $l\le m<a$, $x_m=v$ for $a\le m\le d$,
$x_m>v$ for $d<m\le r$.  The basic idea is to work with the five
inner parts of the array
\begin{equation}
\begin{tabular}{llllrrrr}
\hline
\multicolumn{1}{|c|}{$x<v$} &
\multicolumn{1}{|c|}{$x=v$} &
\multicolumn{1}{|c|}{$x<v$} &
\multicolumn{2}{|c|}{?} &
\multicolumn{1}{|c|}{$x>v$} &
\multicolumn{1}{|c|}{$x=v$} &
\multicolumn{1}{|c|}{$x>v$}\\
\hline
\vphantom{$1^{{2^3}^4}$} % Need more vertical space!
$l$ & $\bar l$ & $p$ & $i$ & $j$ & $q$ & $\bar r$ & $r$\\
\end{tabular}
\label{ternbeg}
\end{equation}
until the middle part is empty or just contains an element equal to the
pivot
\begin{equation}
\begin{tabular}{llrclrr}
\hline
\multicolumn{1}{|c|}{$x=v$} &
\multicolumn{2}{|c|}{$x<v$} &
\multicolumn{1}{|c|}{$x=v$} &
\multicolumn{2}{|c|}{$x>v$} &
\multicolumn{1}{|c|}{$x=v$} \\
\hline
\vphantom{$1^{{2^3}^4}$} % Need more vertical space!
$\bar l$ & $p$ & $j$ & & $i$ & $q$ & $\bar r$ \\
\end{tabular}%\ ,
\label{ternmid}
\end{equation}
(i.e., $j=i-1$ or $j=i-2$),
then swap the ends into the middle for the final arrangement
\begin{equation}
\begin{tabular}{llrr}
\hline
\multicolumn{1}{|c|}{$x<v$} &
\multicolumn{2}{|c|}{$x=v$} &
\multicolumn{1}{|c|}{$x>v$} \\
\hline
\vphantom{$1^{{2^3}^4}$} % Need more vertical space!
$\bar l$ & $a$ & $d$ & $\bar r$\\
\end{tabular}\ .
\label{ternend}
\end{equation}
\begin{description}
\item[A1.] [Initialize.]
Set $v:=x_k$ and exchange $x_l\leftrightarrow x_k$.
Set $i:=\bar l:=l$, $p:=l+1$, $j:=\bar r:=r$, $q:=r-1$.
If $v<x_r$, set $\bar r:=r-1$.  If $v>x_r$, exchange
$x_l\leftrightarrow x_r$ and set $\bar l:=l+1$.
\item[A2.] [Increase $i$ until $x_i\ge v$.]
Increase $i$ by $1$; then if $x_i<v$, repeat this step.
\item[A3.] [Decrease $j$ until $x_j\le v$.]
Decrease $j$ by $1$; then if $x_j>v$, repeat this step.
\item[A4.] [Exchange.]
(Here $x_j\le v\le x_i$.)
If $i<j$, exchange $x_i\leftrightarrow x_j$; then
if $x_i=v$, exchange $x_i\leftrightarrow x_p$ and increase $p$ by $1$;
if $x_j=v$, exchange $x_j\leftrightarrow x_q$ and decrease $q$ by $1$;
return to A2.  If $i=j$ (so that $x_i=x_j=v$), increase $i$ by $1$ and
decrease $j$ by $1$.
\item[A5.] [Cleanup.]
Exchange $x[\bar l\colon p-1]\leftrightarrow x[p\colon j]$ and
$x[i\colon q]\leftrightarrow x[q+1\colon \bar r]$.
Finally, set $a:=\bar l+j-p+1$ and $d:=\bar r-q+i-1$.
\end{description}

Step A1 ensures that $x_l\le v\le x_r$, so steps A2 and A3 don't need
to test whether $i\le j$; thus their loops can run faster than those
in the schemes of
\cite{bemc:esf,bese:fas} and \cite[Ex.\ 5.2.2--41]{knu:acpI3} (which
do need such tests, since, e.g., there may be no element $x_i>v$).
%
%   *** SUBSECTION 6.2 ***
\subsection{Preparing for quintary partitions}
\label{ss:prepquint}
At Step 1, $r-l+1$ replaces $n$ in finding $s$ and $g$.
At Step 2, it is convenient to place the sample in the initial part of
$x[l\colon r]$ by exchanging $x_i\leftrightarrow x_{i+{\rm rand}(r-i)}$
for $l\le i\le r_s:=l+s-1$, where ${\rm rand}(r-i)$ denotes a random
integer, uniformly distributed between $0$ and $r-i$.

Step 3 uses $k_u:=\max\{\lceil l-1+is/m-g\rceil,l\}$ and
$k_v:=\min\{\lceil l-1+is/m+g\rceil,r_s\}$ with $i:=k-l+1$ and $m:=r-l+1$ for
the recursive calls.  If {\sc Select}$(x,l,r_s,k_u,k_u^-,k_u^+)$ returns
$k_u^+\ge k_v$, we have $v:=u:=x_{k_u}$, so we only set $k_v^-:=k_v$,
$k_v^+:=k_u^+$ and reset $k_u^+:=k_v-1$.  Otherwise the second call
{\sc Select}$(x,k_u^++1,r_s,k_v,k_v^-,k_v^+)$ produces $v:=x_{k_v}$.

After $u$ and $v$ have been found, our array looks as follows
\begin{equation}
\begin{tabular}{llrclrrlr}
\hline
\multicolumn{1}{|c|}{$x<u$} &
\multicolumn{2}{|c|}{$x=u$} &
\multicolumn{1}{|c|}{$u<x<v$} &
\multicolumn{2}{|c|}{$x=v$} &
\multicolumn{1}{|c|}{$x>v$} &
\multicolumn{2}{|c|}{?}\\
\hline
\vphantom{$1^{{2^3}^4}$} % Need more vertical space!
$l$ & $k_u^-$ & $k_u^+$ & & $k_v^-$ & $k_v^+$ & $r_s$ & & $r$\\
\end{tabular}\ .
\label{partrec}
\end{equation}
Setting $\bar l:=k_u^-$, $\bar p:=k_u^++1$, $\bar r:=r-r_s+k_v^+$,
$\bar q:=\bar r-k_v^++k_v^--1$, we exchange
$x[k_v^++1\colon r_s]\leftrightarrow x[r_s+1\colon r]$ and then
$x[k_v^-\colon k_v^+]\leftrightarrow x[k_v^++1\colon \bar r]$ to get
the arrangement
\begin{equation}
\begin{tabular}{llllrrr}
\hline
\multicolumn{1}{|c|}{$x<u$} &
\multicolumn{1}{|c|}{$x=u$} &
\multicolumn{1}{|c|}{$u<x<v$} &
\multicolumn{2}{|c|}{?} &
\multicolumn{1}{|c|}{$x=v$} &
\multicolumn{1}{|c|}{$x>v$}\\
\hline
\vphantom{$1^{{2^3}^4}$} % Need more vertical space!
$l$ & $\bar l$ & $\bar p$ & & $\bar q$ & $\bar r$ & $r$\\
\end{tabular}\ .
\label{partini}
\end{equation}
The third part above is missing precisely when $u=v$; in this case
\eqref{partini} reduces to \eqref{ternbeg} with initial $p:=\bar p$,
$q:=\bar q$, $i:=p-1$ and $j:=q+1$.  Hence the case of $u=v$ is handled
via the ternary partitioning scheme of \S\ref{ss:tern}, with step A1
omitted.
%
%   *** SUBSECTION 6.3 ***
\subsection{Quintary partitions}
\label{ss:quint}
For the case of $k<\lfloor(r+l)/2\rfloor$ and $u<v$, Step 4 may use the
following {\em quintary\/} scheme to partition $x[l\colon r]$ into five
blocks, with $x_m<u$ for $l\le m<a$, $x_m=u$ for $a\le m<b$, $u<x_m<v$
for $b\le m\le c$, $x_m=v$ for $c<m\le d$, $x_m>v$ for $d<m\le r$.
The basic idea is to work with the six-part array
stemming from \eqref{partini}
\begin{equation}
\begin{tabular}{llllrrr}
\hline
\multicolumn{1}{|c|}{$x=u$} &
\multicolumn{1}{|c|}{$u<x<v$} &
\multicolumn{1}{|c|}{$x<u$} &
\multicolumn{2}{|c|}{?} &
\multicolumn{1}{|c|}{$x>v$} &
\multicolumn{1}{|c|}{$x=v$} \\
\hline
\vphantom{$1^{{2^3}^4}$} % Need more vertical space!
$\bar l$ & $\bar p$ & $p$ & $i$ & $j$ & $q$ & $\bar r$ \\
\end{tabular}
\label{quintbegleft}
\end{equation}
until $i$ and $j$ cross
\begin{equation}
\begin{tabular}{lllrlrr}
\hline
\multicolumn{1}{|c|}{$x=u$} &
\multicolumn{1}{|c|}{$u<x<v$} &
\multicolumn{2}{|c|}{$x<u$} &
\multicolumn{2}{|c|}{$x>v$} &
\multicolumn{1}{|c|}{$x=v$}\\
\hline
\vphantom{$1^{{2^3}^4}$} % Need more vertical space!
$\bar l$ & $\bar p$ & $p$ & $j$ & $i$ & $q$ & $\bar r$ \\
\end{tabular}\ ;
\label{quintcrossleft}
\end{equation}
we may then swap the second part with the third one to bring it into
the middle
\begin{equation}
\begin{tabular}{lllrrrr}
\hline
\multicolumn{1}{|c|}{$x=u$} &
\multicolumn{1}{|c|}{$x<u$} &
\multicolumn{2}{|c|}{$u<x<v$} &
\multicolumn{2}{|c|}{$x>v$} &
\multicolumn{1}{|c|}{$x=v$} \\
\hline
\vphantom{$1^{{2^3}^4}$} % Need more vertical space!
$\bar l$ & $\bar p$ & $b$ & $c$ & $i$ & $q$ & $\bar r$ \\
\end{tabular}
\label{quintswapleft}
\end{equation}
and finally swap the extreme parts with their neighbors to get the
desired arrangement
\begin{equation}
\begin{tabular}{lllrrr}
\hline
\multicolumn{1}{|c|}{$x<u$} &
\multicolumn{1}{|c|}{$x=u$} &
\multicolumn{2}{|c|}{$u<x<v$} &
\multicolumn{1}{|c|}{$x=v$} &
\multicolumn{1}{|c|}{$x>v$} \\
\hline
\vphantom{$1^{{2^3}^4}$} % Need more vertical space!
$\bar l$ & $a$ & $b$ & $c$ & $d$ & $\bar r$ \\
\end{tabular}\ .
\label{quintend}
\end{equation}
\begin{description}
\item[B1.] [Initialize.]
Set $p:=k_v^-$, $q:=\bar q$, $i:=p-1$ and $j:=q+1$.
\item[B2.] [Increase $i$ until $x_i\ge v$.]
Increase $i$ by $1$.  If $x_i\ge v$, go to B3.
If $x_i<u$, repeat this step.  (At this point, $u\le x_i<v$.)
If $x_i=u$, exchange $x_i\leftrightarrow x_p$ and
$x_p\leftrightarrow x_{\bar p}$ and increase $\bar p$ by $1$;
otherwise exchange $x_i\leftrightarrow x_p$.
Increase $p$ by $1$ and repeat this step.
\item[B3.] [Decrease $j$ until $x_j<v$.]
Decrease $j$ by $1$.  If $x_j>v$, repeat this step.
If $x_j=v$, exchange $x_j\leftrightarrow x_q$, decrease $q$ by $1$
and repeat this step.
\item[B4.] [Exchange.]
If $i\ge j$, go to B5.  Exchange $x_i\leftrightarrow x_j$.
If $x_i=u$, exchange $x_i\leftrightarrow x_p$ and
$x_p\leftrightarrow x_{\bar p}$ and increase $\bar p$ and $p$ by $1$.
Otherwise if $x_i>u$, exchange $x_i\leftrightarrow x_p$ and increase
$p$ by $1$.
If $x_j=v$, exchange $x_j\leftrightarrow x_q$ and decrease $q$ by $1$.
Return to B2.
\item[B5.] [Cleanup.]
Set $a:=\bar l+j-p+1$, $b:=\bar p-p+i$, $c:=j$ and
$d:=\bar r-q+i-1$.
Swap $x[\bar p\colon p-1]\leftrightarrow x[p\colon j]$,
$x[\bar l\colon\bar p-1]\leftrightarrow x[\bar p\colon b-1]$,
and finally $x[i\colon q]\leftrightarrow x[q+1\colon\bar r]$.
\end{description}

For the case of $k\ge\lfloor(r+l)/2\rfloor$ and $u<v$, Step 4 may use
the following quintary scheme, which is a symmetric version of the
preceding one obtained by replacing
\eqref{quintbegleft}--\eqref{quintswapleft} by
\begin{equation}
\begin{tabular}{lllrrrr}
\hline
\multicolumn{1}{|c|}{$x=u$} &
\multicolumn{1}{|c|}{$x<u$} &
\multicolumn{2}{|c|}{?} &
\multicolumn{1}{|c|}{$x>v$} &
\multicolumn{1}{|c|}{$u<x<v$} &
\multicolumn{1}{|c|}{$x=v$} \\
\hline
\vphantom{$1^{{2^3}^4}$} % Need more vertical space!
$\bar l$ & $p$ & $i$ & $j$ & $q$ & $\bar q$ & $\bar r$ \\
\end{tabular}\ ,
\label{quintbegright}
\end{equation}
\begin{equation}
\begin{tabular}{llrlrrr}
\hline
\multicolumn{1}{|c|}{$x=u$} &
\multicolumn{2}{|c|}{$x<u$} &
\multicolumn{2}{|c|}{$x>v$} &
\multicolumn{1}{|c|}{$u<x<v$} &
\multicolumn{1}{|c|}{$x=v$}\\
\hline
\vphantom{$1^{{2^3}^4}$} % Need more vertical space!
$\bar l$ & $p$ & $j$ & $i$ & $q$ & $\bar q$ & $\bar r$ \\
\end{tabular}\ ,
\label{quintcrossright}
\end{equation}
\begin{equation}
\begin{tabular}{lllrrr}
\hline
\multicolumn{1}{|c|}{$x<u$} &
\multicolumn{1}{|c|}{$x=u$} &
\multicolumn{2}{|c|}{$u<x<v$} &
\multicolumn{1}{|c|}{$x>v$} &
\multicolumn{1}{|c|}{$x=v$}\\
\hline
\vphantom{$1^{{2^3}^4}$} % Need more vertical space!
$\bar l$ & $a$ & $b$ & $c$ & $\bar q$ & $\bar r$ \\
\end{tabular}\ .
\label{quintswapright}
\end{equation}
\begin{description}
\item[C1.] [Initialize.]
Set $p:=\bar p$, $q:=\bar q-k_v^-+k_u^++1$, $i:=p-1$ and $j:=q+1$,
and swap
$x[\bar p\colon k_v^--1]\leftrightarrow x[k_v^-\colon\bar q]$.
\item[C2.] [Increase $i$ until $x_i>u$.]
Increase $i$ by $1$.  If $x_i<u$, repeat this step.
If $x_i=u$, exchange $x_i\leftrightarrow x_p$, increase $p$ by $1$
and repeat this step.
\item[C3.] [Decrease $j$ until $x_j\le u$.]
Decrease $j$ by $1$.  If $x_j\le u$, go to C4.
If $x_j>v$, repeat this step.  (At this point, $u<x_j\le v$.)
If $x_j=v$, exchange $x_j\leftrightarrow x_q$ and
$x_q\leftrightarrow x_{\bar q}$ and decrease $\bar q$ by $1$;
otherwise exchange $x_j\leftrightarrow x_q$.
Decrease $q$ by $1$ and repeat this step.
\item[C4.] [Exchange.]
If $i\ge j$, go to C5.  Exchange $x_i\leftrightarrow x_j$.
If $x_i=u$, exchange $x_i\leftrightarrow x_p$ and increase $p$ by $1$.
If $x_j=v$, exchange $x_j\leftrightarrow x_q$ and
$x_q\leftrightarrow x_{\bar q}$ and decrease $\bar q$ and $q$ by $1$.
Otherwise if $x_j>v$, exchange $x_j\leftrightarrow x_q$ and decrease
$q$ by $1$.  Return to C2.
\item[C5.] [Cleanup.]
Set $a:=\bar l+j-p+1$, $b:=i$, $c:=\bar q-q+j$ and
$d:=\bar r-q+i-1$.
Swap $x[i\colon q]\leftrightarrow x[q+1\colon\bar q]$,
$x[c+1\colon\bar q]\leftrightarrow x[\bar q+1\colon\bar r]$,
and finally $x[\bar l\colon p-1]\leftrightarrow x[p\colon j]$.
\end{description}

To make \eqref{ternend} and \eqref{quintend} compatible, the ternary
scheme may set $b:=d+1$, $c:=a-1$.  After partitioning $l$ and $r$
are updated by setting $l:=b$ if $a\le k$, then $l:=d+1$ if $c<k$;
$r:=c$ if $k\le d$, then $r:=a-1$ if $k<b$.  If $l\ge r$,
{\sc Select} may return $k_-:=k_+:=k$ if $l=r$, $k_-:=r+1$ and
$k_+:=l-1$ if $l>r$.  Otherwise, instead of calling {\sc Select}
recursively, Step 6 may jump back to Step 1, or Step 0 if sSelect
is used (cf.\ \S\ref{ss:subfile}).

A simple version of sSelect is
obtained if Steps 2 and 3 choose $u:=v:=x_k$ when
$r-l+1\le n_{\rm cut}$ (this choice of \cite{flri:asf} works well in
practice, but more sophisticated pivots could be tried); then the
ternary partitioning code can be used by sSelect as well.

In fact steps A5, B5 and C5 may also share code: resetting
$\bar q:=q$ and $\bar p:=p$ for A5, $\bar q:=q$ for B5, and
$\bar p:=p$ for C5, we may swap
$x[\bar p\colon p-1]\leftrightarrow x[p\colon j]$ if $p>\bar p$,
$x[\bar l\colon\bar p-1]\leftrightarrow x[\bar p\colon\bar p-p+j]$,
$x[i\colon q]\leftrightarrow x[q+1\colon\bar q]$ if $q<\bar q$,
$x[i+\bar q-q\colon\bar q]\leftrightarrow x[\bar q+1\colon\bar r]$.

Even when outcomes of previous comparisons are utilized, our schemes
still involve two extraneous comparisons (scheme A only one when $i=j$
at A4).  Consider, therefore, the following alternative to scheme B,
also based on the arrangements \eqref{quintbegleft}--\eqref{quintend}.
\begin{description}
\item[D1.] [Initialize.]
Set $p:=k_v^-$, $q:=\bar q$, $i:=p$, $j:=q$.
\item[D2.] [Increase $i$ until $x_i\ge v$.]
If $i>j$ or $x_i\ge v$, go to D3.
If $x_i=u$, exchange $x_i\leftrightarrow x_p$ and
$x_p\leftrightarrow x_{\bar p}$, and increase $p$ and $\bar p$ by $1$;
otherwise if $x_i>u$, exchange $x_i\leftrightarrow x_p$ and increase
$p$ by $1$.  Increase $i$ by $1$ and repeat this step.
\item[D3.] [Decrease $j$ until $x_j<v$.]
If $i>j$ or $x_j<v$, go to D4.
If $x_j=v$, exchange $x_j\leftrightarrow x_q$ and decrease $q$ by $1$.
Decrease $j$ by $1$ and repeat this step.
\item[D4.] [Exchange.]
If $i\ge j$, go to D5.  Exchange $x_i\leftrightarrow x_j$.
If $x_i=u$, exchange $x_i\leftrightarrow x_p$ and
$x_p\leftrightarrow x_{\bar p}$ and increase $\bar p$ and $p$ by $1$.
Otherwise if $x_i>u$, exchange $x_i\leftrightarrow x_p$ and increase
$p$ by $1$.
If $x_j=v$, exchange $x_j\leftrightarrow x_q$ and decrease $q$ by $1$.
Increase $i$ by $1$, decrease $j$ by $1$, and return to D2.
\item[D5.] [Cleanup.]
Set $a:=\bar l+j-p+1$, $b:=\bar p-p+i$, $c:=j$ and
$d:=\bar r-q+i-1$.
Swap $x[\bar p\colon p-1]\leftrightarrow x[p\colon j]$,
$x[\bar l\colon\bar p-1]\leftrightarrow x[\bar p\colon b-1]$,
and finally $x[i\colon q]\leftrightarrow x[q+1\colon\bar r]$.
\end{description}
Relative to scheme B (which makes $r-r_s+2$ comparisons to $v$), scheme
D saves two $v$-comparisons at the expense of $r-r_s+2$ comparisons of
$i$ vs.\ $j$.  Since $r-r_s\gg2$ for usual choices of $n_{\rm cut}$,
scheme B is faster than D unless the cost of key comparisons is
extremely large.  Scheme C compares in the same way with a symmetric
variant of D.  The situation with scheme A is similar, even
for the small partitions produced in sSelect: although the schemes of
\cite{bemc:esf,bese:fas} can save two $v$-comparisons, such savings
are insignificant when relatively few small partitions occur
(cf.\ \S\ref{ss:result}).
%
%   *** SUBSECTION 6.4 ***
\subsection{Poor man's partitions}
\label{ss:poor}
We now consider a {\em poor man's\/} version of {\sc Select},
called {\sc pmSelect}, which employs less refined but hopefully
faster partitioning.  This version works with $x[l\colon r]$ such that
$x_i\le x_l$ for $i=1\colon l-1$, $x_r\le x_i$ for $i=r+1\colon n$,
and its task is standard: given $x[l\colon r]$ and $l\le k\le r$,
{\sc pmSelect}$(x,l,r,k)$ permutes $x[l\colon r]$ so that
$x_i\le x_k$ for all $l\le i<k$, and $x_k\le x_i$ for all $k<i\le r$;
the initial call is {\sc pmSelect}$(x,1,n,k)$.  We start with binary
partitions.

For a given pivot $v:=x_k$ from the array $x[l\colon r]$, the
following {\em binary\/} scheme partitions the array into three blocks,
with $x_m\le v$ for $l\le m<a$, $x_m=v$ for $a\le m\le d$,
$v\le x_m$ for $d<m\le r$; usually $a=d$ and the middle block is
singleton.
\begin{description}
\item[E1.] [Initialize.]
Set $v:=x_k$ and exchange $x_l\leftrightarrow x_k$.
Set $i:=\hat p:=l$ and $j:=r$.  If $v>x_r$, exchange
$x_l\leftrightarrow x_r$ and set $\hat p:=r$.  (Thus $v=x_{\hat p}$
always.)
\item[E2.] [Increase $i$ until $x_i\ge v$.]
Increase $i$ by $1$; then if $x_i<v$, repeat this step.
\item[E3.] [Decrease $j$ until $x_j\le v$.]
Decrease $j$ by $1$; then if $x_j>v$, repeat this step.
\item[E4.] [Exchange.]
(Here $x_j\le v\le x_i$.)
If $i<j$, exchange $x_i\leftrightarrow x_j$ and return to E2.  If $i=j$
(so that $x_i=x_j=v$), increase $i$ by $1$ and decrease $j$ by $1$.
\item[E5.] [Cleanup.]
If $\hat p\ne r$, exchange $x_{\hat p}\leftrightarrow x_j$, set
$a:=j$ and $d:=i-1$; otherwise exchange $x_i\leftrightarrow x_{\hat p}$,
set $a:=j+1$ and $d:=i$.
\end{description}

The setup of \S\ref{ss:prepquint} changes as follows.
Step 3 calls {\sc pmSelect}$(x,l,r_s,k_u)$ to find $u:=x_{k_u}$, and
then {\sc pmSelect}$(x,k_u+1,r_s,k_v)$ to get $v:=x_{k_v}$, assuming
$k_u<k_v$; then \eqref{partrec} changes to
\begin{equation}
\begin{tabular}{lcccrlr}
\hline
\multicolumn{1}{|c|}{$x\le u$} &
\multicolumn{1}{|c|}{$u$} &
\multicolumn{1}{|c|}{$u\le x\le v$} &
\multicolumn{1}{|c|}{$v$} &
\multicolumn{1}{|c|}{$x\ge v$} &
\multicolumn{2}{|c|}{?}\\
\hline
\vphantom{$1^{{2^3}^4}$} % Need more vertical space!
$l$ & $k_u$ & & $k_v$ & $r_s$ & & $r$\\
\end{tabular}\ .
\label{partrecpm}
\end{equation}
Setting $\bar l:=k_u$, $p:=k_v$, $\bar r:=r-r_s+p$, we exchange
$x[p+1\colon r_s]\leftrightarrow x[r_s+1\colon r]$ and then
$x_p\leftrightarrow x_{\bar r}$ to get the arrangement
\begin{equation}
\begin{tabular}{lcclrclr}
\hline
\multicolumn{1}{|c|}{$x\le u$} &
\multicolumn{1}{|c|}{$u$} &
\multicolumn{1}{|c|}{$u\le x\le v$} &
\multicolumn{2}{|c|}{?} &
\multicolumn{1}{|c|}{$v$} &
\multicolumn{2}{|c|}{$x\ge v$}\\
\hline
\vphantom{$1^{{2^3}^4}$} % Need more vertical space!
$l$ & $\bar l$ & & $p$ & & $\bar r$ & & $r$\\
\end{tabular}\ .
\label{partinipm}
\end{equation}
If $u=v$, setting $i:=p-1$ and $j:=\bar r$, we may use scheme E with
step E1 omitted and E5 replaced by A5 with $q:=\bar r-1$; i.e., E5 now
reads
\begin{description}
\item[E5.] [Cleanup.]
Set $a:=\bar l+j-p+1$ and $d:=i$.
Swap $x[\bar l\colon p-1]\leftrightarrow x[p\colon j]$ and
$x_d\leftrightarrow x_{\bar r}$.
\end{description}

For the case of $k<\lfloor(r+l)/2\rfloor$ and $u<v$, we may use the
following quintary scheme, obtained by replacing
\eqref{quintbegleft}--\eqref{quintswapleft} with the arrangements
\begin{equation}
\begin{tabular}{lcllrcl}
\hline
\multicolumn{1}{|c|}{$u$} &
\multicolumn{1}{|c|}{$u\le x\le v$} &
\multicolumn{1}{|c|}{$x\le u$} &
\multicolumn{2}{|c|}{?} &
\multicolumn{1}{|c|}{$x\ge v$} &
\multicolumn{1}{|c|}{$v$} \\
\hline
\vphantom{$1^{{2^3}^4}$} % Need more vertical space!
$\bar l$ & & $p$ & $i$ & $j$ & & $\bar r$ \\
\end{tabular}\ ,
\label{quintbegleftpm}
\end{equation}
\begin{equation}
\begin{tabular}{lclrlc}
\hline
\multicolumn{1}{|c|}{$u$} &
\multicolumn{1}{|c|}{$u\le x\le v$} &
\multicolumn{2}{|c|}{$x\le u$} &
\multicolumn{1}{|c|}{$x\ge v$} &
\multicolumn{1}{|c|}{$v$} \\
\hline
\vphantom{$1^{{2^3}^4}$} % Need more vertical space!
$\bar l$ & & $p$ & $j$ & $i$ & $\bar r$\\
\end{tabular}\ ,
\label{quintcrossleftpm}
\end{equation}
\begin{equation}
\begin{tabular}{lrrcr}
\hline
\multicolumn{1}{|c|}{$u$} &
\multicolumn{1}{|c|}{$x\le u$} &
\multicolumn{1}{|c|}{$u\le x\le v$} &
\multicolumn{1}{|c|}{$v$} &
\multicolumn{1}{|c|}{$x\ge v$}\\
\hline
\vphantom{$1^{{2^3}^4}$} % Need more vertical space!
$\bar l$ & $a$ & $j$ & $i$ & $\bar r$\\
\end{tabular}\ ,
\label{quintswapleftpm}
\end{equation}
\begin{equation}
\begin{tabular}{lclrcr}
\hline
\multicolumn{1}{|c|}{$x\le u$} &
\multicolumn{1}{|c|}{$u$} &
\multicolumn{2}{|c|}{$u\le x\le v$} &
\multicolumn{1}{|c|}{$v$} &
\multicolumn{1}{|c|}{$x\ge v$}\\
\hline
\vphantom{$1^{{2^3}^4}$} % Need more vertical space!
$\bar l$ & $a$ & $b$ & $c$ & $d$ & $\bar r$\\
\end{tabular}\ .
\label{quintendpm}
\end{equation}
\begin{description}
\item[F1.] [Initialize.]
Set $i:=p-1$ and $j:=\bar r$.
\item[F2.] [Increase $i$ until $x_i\ge v$.]
Increase $i$ by $1$.  If $x_i\ge v$, go to F3.
If $x_i\le u$, repeat this step.
(Here $u<x_i<v$.)
%(At this point, $u<x_i<v$.)
Exchange $x_i\leftrightarrow x_p$, increase $p$ by $1$ and repeat this
step.
\item[F3.] [Decrease $j$ until $x_j<v$.]
Decrease $j$ by $1$.  If $x_j\ge v$, repeat this step.
\item[F4.] [Exchange.]
If $i\ge j$, go to F5.  Exchange $x_i\leftrightarrow x_j$.
If $x_i>u$, exchange $x_i\leftrightarrow x_p$ and increase
$p$ by $1$.  Return to F2.
\item[F5.] [Cleanup.]
Set $a:=\bar l+i-p$, $b:=a+1$, $c:=j$ and $d:=j+1$.
Swap $x[\bar l+1\colon p-1]\leftrightarrow x[p\colon j]$,
$x_{\bar l}\leftrightarrow x_a$ and finally
$x_d\leftrightarrow x_{\bar r}$.
\end{description}

In fact scheme F may produce $j<i-1$ on the first pass if
$x_m\ge v$ for all $m\ge k_v-1$.

For the case of $k\ge\lfloor(r+l)/2\rfloor$ and $u<v$, we may use the
following quintary scheme, which is a symmetric version of scheme F
obtained by replacing \eqref{quintbegleftpm}--\eqref{quintswapleftpm} by
\begin{equation}
\begin{tabular}{cclrrcc}
\hline
\multicolumn{1}{|c|}{$u$} &
\multicolumn{1}{|c|}{$x\le u$} &
\multicolumn{2}{|c|}{?} &
\multicolumn{1}{|c|}{$x\ge v$} &
\multicolumn{1}{|c|}{$u\le x\le v$} &
\multicolumn{1}{|c|}{$v$} \\
\hline
\vphantom{$1^{{2^3}^4}$} % Need more vertical space!
$\bar l$ & & $i$ & $j$ & $q$ & & $\bar r$ \\
\end{tabular}\ ,
\label{quintbegrightpm}
\end{equation}
\begin{equation}
\begin{tabular}{crlrccr}
\hline
\multicolumn{1}{|c|}{$u$} &
\multicolumn{1}{|c|}{$x\le u$} &
\multicolumn{2}{|c|}{$x\ge v$} &
\multicolumn{1}{|c|}{$u\le x\le v$} &
\multicolumn{1}{|c|}{$v$}\\
\hline
\vphantom{$1^{{2^3}^4}$} % Need more vertical space!
$\bar l$ & $j$ & $i$ & $q$ & & $\bar r$\\
\end{tabular}\ ,
\label{quintcrossrightpm}
\end{equation}
\begin{equation}
\begin{tabular}{lcllcr}
\hline
\multicolumn{1}{|c|}{$x\le u$} &
\multicolumn{1}{|c|}{$u$} &
\multicolumn{1}{|c|}{$u\le x\le v$} &
\multicolumn{1}{|c|}{$x\ge v$} &
\multicolumn{1}{|c|}{$v$}\\
\hline
\vphantom{$1^{{2^3}^4}$} % Need more vertical space!
$\bar l$ & $j$ & $i$ & $d$ & $\bar r$\\
\end{tabular}\ .
\label{quintswaprightpm}
\end{equation}
\begin{description}
\item[G1.] [Initialize.]
Set $q:=r-r_s+\bar l$, $i:=\bar l$, $j:=q+1$, and
swap $x[\bar l+1\colon p-1]\leftrightarrow x[p\colon\bar r-1]$.
\item[G2.] [Increase $i$ until $x_i>u$.]
Increase $i$ by $1$.  If $x_i\le u$, repeat this step.
\item[G3.] [Decrease $j$ until $x_j\le u$.]
Decrease $j$ by $1$.  If $x_j\le u$, go to G4.
If $x_j\ge v$, repeat this step.  (Here $u<x_j<v$.)
Exchange $x_j\leftrightarrow x_q$, decrease $q$ by $1$ and repeat this
step.
\item[G4.] [Exchange.]
If $i\ge j$, go to G5.  Exchange $x_i\leftrightarrow x_j$.
If $x_j<v$, exchange $x_j\leftrightarrow x_q$ and decrease
$q$ by $1$.  Return to G2.
\item[G5.] [Cleanup.]
Set $a:=j$, $b:=a+1$, $d:=\bar r-q+j$ and $c:=d-1$.
Exchange $x_{\bar l}\leftrightarrow x_a$,
$x[i\colon q]\leftrightarrow x[q+1\colon\bar r-1]$,
and finally $x_d\leftrightarrow x_{\bar r}$.
\end{description}

Also scheme G may produce $j<i-1$ on the first pass if
$x_m<v$ for all $m\le q+1$.

Schemes E, F and G are like A, B and C without their equality tests
and associated updates.  When equal elements are absent, the
inequalities in \eqref{partrecpm}--\eqref{quintswaprightpm} are strict,
so E, F and G (although {\em not\/} equivalent to A, B and C) yield
correct partitions for Step 4.  When equal elements occur, the
partitions of E, F and G needn't meet the requirements of Step 4, but
are still usable.  Namely, $r-l$ shrinks when $a$, $b$, $c$ and $d$ are
used for updating $l$ and $r$, and sSelect may employ scheme E instead
of A as in \S\ref{ss:quint}.  In effect, {\sc pmSelect} works like
{\sc Select} in the case of distinct elements, but may require more
comparisons otherwise.  In practice {\sc pmSelect} tends to be slightly
faster (cf.\ \S\ref{ss:result}).
%
%   *** SECTION 7 ***
\section{Experimental results}
\label{s:exp}
%
%   *** SUBSECTION 7.1 ***
\subsection{Implemented algorithms}
\label{ss:impl}
An implementation of {\sc Select} was programmed in Fortran 77 and
run on a notebook PC (Pentium II 400 MHz, 256 MB RAM) under MS
Windows 98.  The input set $X$ was specified as a double precision
array.  For efficiency, the recursion was removed and small arrays with
$n\le n_{\rm cut}$ were handled as if Steps 2 and 3 chose $u:=v:=x_k$;
the resulting version of sSelect (cf.\ \S\S\ref{ss:subfile} and
\ref{ss:quint}) typically
required less than $3.5n$ comparisons.  The choice of \eqref{sgf} was
employed, with the parameters $\alpha=0.5$, $\beta=0.25$ and
$n_{\rm cut}=600$ as proposed in \cite{flri:asf}; future work should
test other sample sizes and parameters.

A similar implementation of {\sc pmSelect} was programmed as described
in \S\ref{ss:poor}.

For comparisons we developed a Fortran 77 implementation of the
{\sc riSelect} algorithm of \cite{val:iss}.  Briefly, {\sc riSelect}
behaves like quickselect using the median of the first, middle and last
elements, these elements being exchanged with randomly chosen ones only
if the file doesn't shrink sufficiently fast.  To ensure $O(n)$ time
in the worst case, {\sc riSelect} may switch to the algorithm of
\cite{blflprrita:tbs}, but this never happened in our experiments.
Apparently {\sc riSelect} represents the state-of-the-art in quickselect
implementations (several other implementations fared worse in out tests).
%
%   *** SUBSECTION 7.2 ***
\subsection{Testing examples}
\label{ss:examp}
We used minor modifications of the input sequences of \cite{val:iss},
defined as follows:
\begin{description}
\item[random]
A random permutation of the integers $1$ through $n$.
\item[onezero]
A random permutation of $\lceil n/2\rceil$ ones and $\lfloor n/2\rfloor$
zeroes.
\item[sorted]
The integers $1$ through $n$ in increasing order.
\item[rotated]
A sorted sequence rotated left once; i.e., $(2,3,\ldots,n,1)$.
\item[organpipe]
The integers $1$ through $n/2$ in increasing order, followed by $n/2$
through $1$ in decreasing order.
\item[m3killer]
Musser's ``median-of-3 killer'' sequence with $n=4j$ and $k=n/2$:
$$
\left(\begin{array}{ccccccccccccc}
1&  2 & 3&  4 & \ldots&  k-2& k-1& k& k+1& \ldots& 2k-2& 2k-1& 2k\\
1& k+1& 3& k+3& \ldots& 2k-3& k-1& 2&  4 & \ldots& 2k-2& 2k-1& 2k
\end{array}\right).
$$
\item[twofaced]
%Musser's ``two-faced'' sequence, obtained by randomly permuting the
Obtained by randomly permuting the
elements of an m3killer sequence in positions $4\lfloor\log_2n\rfloor$
through $n/2-1$ and $n/2+4\lfloor\log_2n\rfloor-1$ through $n-2$.
\end{description}
For each input sequence, its (lower) median element was found.

These input sequences were designed to test the performance of
selection algorithms under a range of conditions.
In particular, the onezero sequences represent inputs containing
many duplicates \cite{sed:qek}.  The rotated and organpipe sequences
are difficult for many implementations of quickselect.  The m3killer
and twofaced sequences are hard for implementations with median-of-3
pivots (their original versions \cite{mus:iss} were modified to become
difficult when the middle element comes from position $k$ instead of
$k+1$).
%
%   *** SUBSECTION 7.3 ***
\subsection{Computational results}
\label{ss:result}
We varied the input size $n$ from $50{,}000$ to $16{,}000{,}000$.  For
the random, onezero and twofaced sequences, for each input size,
20 instances were randomly generated; for the deterministic
sequences, five runs were made to measure the solution time.

The performance of {\sc Select} on randomly generated inputs is
summarized in Table \ref{tab:Selrand},
%
%   *** TABLE 7.1 ***
\begin{table}[t!]
\caption{Performance of {\sc Select} on randomly generated inputs.}
\label{tab:Selrand}
\footnotesize
\begin{center}
\begin{tabular}{lrrrrrrrrrrrrr}
\hline
Sequence &\multicolumn{1}{c}{Size}
&\multicolumn{3}{c}{Time $[{\rm sec}]$%
\vphantom{$1^{2^3}$}} % Need more vertical space!
&\multicolumn{3}{c}{Comparisons $[n]$}
&\multicolumn{1}{c}{$\gamma_{\rm avg}$}
&\multicolumn{1}{c}{$L_{\rm avg}$}
&\multicolumn{1}{c}{$P_{\rm avg}$}
&\multicolumn{1}{c}{$N_{\rm avg}$}
&\multicolumn{1}{c}{$p_{\rm avg}$}
&\multicolumn{1}{c}{$s_{\rm avg}$}\\
&\multicolumn{1}{c}{$n$}
&\multicolumn{1}{c}{avg}&\multicolumn{1}{c}{max}&\multicolumn{1}{c}{min}
&\multicolumn{1}{c}{avg}&\multicolumn{1}{c}{max}&\multicolumn{1}{c}{min}
& &\multicolumn{1}{c}{$[n]$}
&\multicolumn{1}{c}{$[\ln n]$}
&\multicolumn{1}{c}{$[\ln n]$} &
&\multicolumn{1}{c}{$[\%n]$}\\
\hline
%alpha=0.5 beta=0.25 cutoff=600
random     &  50K
& 0.01& 0.06& 0.01& 1.79& 1.84& 1.74& 4.91& 1.21& 0.46& 1.01& 7.40& 4.10\\
           & 100K
& 0.02& 0.06& 0.01& 1.73& 1.77& 1.70& 4.77& 1.15& 0.43& 0.96& 8.03& 3.20\\
           & 500K
& 0.06& 0.11& 0.05& 1.62& 1.63& 1.61& 4.06& 1.08& 0.56& 1.20& 8.00& 1.86\\
           &   1M
& 0.12& 0.17& 0.11& 1.59& 1.60& 1.58& 3.95& 1.06& 0.67& 1.40& 7.95& 1.47\\
           &   2M
& 0.22& 0.22& 0.21& 1.57& 1.58& 1.56& 3.76& 1.04& 0.76& 1.59& 7.90& 1.16\\
           &   4M
& 0.43& 0.44& 0.38& 1.56& 1.56& 1.55& 3.63& 1.03& 0.95& 1.95& 7.29& 0.92\\
           &   8M
& 0.83& 0.88& 0.82& 1.54& 1.55& 1.54& 3.54& 1.03& 0.98& 2.00& 7.41& 0.72\\
           &  16M
& 1.62& 1.65& 1.59& 1.53& 1.54& 1.53& 3.39& 1.02& 1.00& 2.05& 7.77& 0.57\\
onezero    &  50K
& 0.01& 0.06& 0.01& 1.52& 1.52& 1.50& 0.25& 1.02& 0.28& 0.27& 1.17& 3.41\\
           & 100K
& 0.02& 0.06& 0.01& 1.51& 1.51& 1.50& 0.24& 1.01& 0.26& 0.25& 1.24& 2.72\\
           & 500K
& 0.07& 0.11& 0.05& 1.51& 1.51& 1.51& 0.26& 1.01& 0.23& 0.23& 1.15& 1.61\\
           &   1M
& 0.13& 0.17& 0.11& 1.51& 1.51& 1.51& 0.26& 1.01& 0.22& 0.22& 1.15& 1.29\\
           &   2M
& 0.27& 0.28& 0.22& 1.51& 1.51& 1.50& 0.26& 1.01& 0.28& 0.27& 1.09& 1.03\\
           &   4M
& 0.54& 0.55& 0.49& 1.50& 1.50& 1.50& 0.26& 1.00& 0.33& 0.26& 1.16& 0.83\\
           &   8M
& 1.02& 1.05& 0.98& 1.50& 1.50& 1.50& 0.26& 1.00& 0.38& 0.25& 1.10& 0.66\\
           &  16M
& 2.04& 2.09& 2.03& 1.50& 1.50& 1.50& 0.26& 1.00& 0.36& 0.24& 1.13& 0.53\\
twofaced   &  50K
& 0.02& 0.06& 0.01& 1.81& 1.84& 1.76& 5.11& 1.21& 0.46& 1.02& 7.78& 4.13\\
           & 100K
& 0.01& 0.06& 0.01& 1.73& 1.77& 1.71& 4.81& 1.16& 0.44& 0.96& 8.01& 3.20\\
           & 500K
& 0.07& 0.11& 0.05& 1.62& 1.63& 1.59& 4.10& 1.08& 0.56& 1.20& 8.15& 1.86\\
           &   1M
& 0.11& 0.16& 0.10& 1.59& 1.60& 1.58& 3.89& 1.06& 0.64& 1.36& 7.82& 1.47\\
           &   2M
& 0.22& 0.27& 0.22& 1.57& 1.58& 1.56& 3.63& 1.04& 0.75& 1.58& 7.63& 1.16\\
           &   4M
& 0.42& 0.44& 0.38& 1.56& 1.56& 1.55& 3.57& 1.03& 0.96& 1.97& 7.29& 0.92\\
           &   8M
& 0.83& 0.88& 0.82& 1.54& 1.55& 1.54& 3.50& 1.03& 0.97& 2.00& 7.43& 0.72\\
           &  16M
& 1.62& 1.65& 1.59& 1.53& 1.54& 1.53& 3.40& 1.02& 1.00& 2.03& 7.57& 0.57\\
\hline
\end{tabular}
\end{center}
\end{table}
where the average, maximum and minimum solution times are in seconds,
and the comparison counts are in multiples of $n$; e.g., column six
gives $C_{\rm avg}/n$, where $C_{\rm avg}$ is the average number of
comparisons made over all instances.  Thus
$\gamma_{\rm avg}:=(C_{\rm avg}-1.5n)/f(n)$ estimates the constant
$\gamma$ in the bound \eqref{CnkFR}; moreover, we have
$C_{\rm avg}\approx1.5L_{\rm avg}$, where $L_{\rm avg}$ is the average
sum of sizes of partitioned arrays.  Further,
$P_{\rm avg}$ is the average number of {\sc Select} partitions, whereas
$N_{\rm avg}$ is the average number of calls to sSelect and
$p_{\rm avg}$ is the average number of sSelect partitions per call;
both $P_{\rm avg}$ and $N_{\rm avg}$ grow slowly with $\ln n$.
Finally, $s_{\rm avg}$ is the average sum of sample sizes;
$s_{\rm avg}/f(n)$ drops from $0.68$ for $n=50{\rm K}$ to $0.56$ for
$n=16{\rm M}$ on the random and twofaced inputs, and from $0.57$ to
$0.52$ on the onezero inputs, whereas the initial
$s/f(n)\approx\alpha=0.5$.
The average solution times grow linearly with $n$ (except for small
inputs whose solution times couldn't be measured accurately), and the
differences between maximum and minimum times are fairly small (and also
partly due to the operating system).  Except for the smallest inputs,
the maximum and minimum numbers of comparisons are quite close, and
$C_{\rm avg}$ nicely approaches the theoretical lower bound of $1.5n$;
this is reflected in the values of $\gamma_{\rm avg}$.  Note that
the results for the random and twofaced sequences are almost identical,
whereas the onezero inputs only highlight the efficiency of our
partitioning.

Table \ref{tab:Seldet} exhibits similar features of {\sc Select} on
the deterministic inputs.
%
%   *** TABLE 7.2 ***
\begin{table}[t!]
\caption{Performance of {\sc Select} on deterministic inputs.}
\label{tab:Seldet}
\footnotesize
\begin{center}
\begin{tabular}{lrrrrcrrrrrr}
\hline
Sequence &\multicolumn{1}{c}{Size}
&\multicolumn{3}{c}{Time $[{\rm sec}]$%
\vphantom{$1^{2^3}$}} % Need more vertical space!
&\multicolumn{1}{c}{Comparisons}
&\multicolumn{1}{c}{$\gamma_{\rm avg}$}
&\multicolumn{1}{c}{$L_{\rm avg}$}
&\multicolumn{1}{c}{$P_{\rm avg}$}
&\multicolumn{1}{c}{$N_{\rm avg}$}
&\multicolumn{1}{c}{$p_{\rm avg}$}
&\multicolumn{1}{c}{$s_{\rm avg}$}\\
&\multicolumn{1}{c}{$n$}
&\multicolumn{1}{c}{avg}&\multicolumn{1}{c}{max}&\multicolumn{1}{c}{min}
&\multicolumn{1}{c}{$[n]$}
& &\multicolumn{1}{c}{$[n]$}
&\multicolumn{1}{c}{$[\ln n]$}
&\multicolumn{1}{c}{$[\ln n]$} &
&\multicolumn{1}{c}{$[\%n]$}\\
\hline
%alpha=0.5 beta=0.25 cutoff=600
sorted     &  50K
& 0.02& 0.06& 0.01& 1.79& 4.91& 1.23& 0.46& 1.02& 8.36& 4.12\\
           & 100K
& 0.01& 0.01& 0.01& 1.73& 4.69& 1.16& 0.43& 0.96& 8.55& 3.21\\
           & 500K
& 0.04& 0.06& 0.01& 1.60& 3.33& 1.07& 0.61& 1.30& 7.71& 1.86\\
           &   1M
& 0.09& 0.11& 0.05& 1.57& 3.07& 1.06& 0.65& 1.38& 6.58& 1.47\\
           &   2M
& 0.14& 0.17& 0.11& 1.56& 2.99& 1.04& 0.76& 1.59& 7.57& 1.15\\
           &   4M
& 0.28& 0.28& 0.28& 1.55& 3.02& 1.03& 0.99& 2.04& 8.06& 0.92\\
           &   8M
& 0.55& 0.55& 0.55& 1.54& 3.12& 1.03& 1.01& 2.01& 7.13& 0.72\\
           &  16M
& 1.05& 1.10& 1.04& 1.53& 3.20& 1.02& 1.02& 2.11& 7.46& 0.57\\
rotated    &  50K
& 0.01& 0.06& 0.01& 1.80& 4.92& 1.23& 0.46& 1.02& 8.55& 4.12\\
           & 100K
& 0.01& 0.01& 0.01& 1.73& 4.69& 1.16& 0.43& 0.96& 8.55& 3.21\\
           & 500K
& 0.02& 0.06& 0.01& 1.60& 3.33& 1.07& 0.61& 1.30& 7.82& 1.86\\
           &   1M
& 0.09& 0.11& 0.05& 1.57& 3.08& 1.06& 0.65& 1.38& 6.74& 1.47\\
           &   2M
& 0.15& 0.17& 0.11& 1.56& 2.99& 1.04& 0.76& 1.59& 7.13& 1.15\\
           &   4M
& 0.28& 0.28& 0.27& 1.55& 3.02& 1.03& 0.99& 2.04& 7.71& 0.92\\
           &   8M
& 0.55& 0.55& 0.54& 1.54& 3.12& 1.03& 1.01& 2.01& 7.19& 0.72\\
           &  16M
& 1.05& 1.10& 1.04& 1.53& 3.20& 1.02& 1.02& 2.11& 7.46& 0.57\\
organpipe  &  50K
& 0.01& 0.05& 0.01& 1.83& 5.43& 1.22& 0.46& 1.02& 8.55& 4.11\\
           & 100K
& 0.03& 0.06& 0.01& 1.74& 4.99& 1.16& 0.43& 0.96& 6.64& 3.19\\
           & 500K
& 0.04& 0.06& 0.01& 1.62& 3.97& 1.07& 0.61& 1.30& 6.65& 1.87\\
           &   1M
& 0.11& 0.11& 0.11& 1.59& 3.77& 1.06& 0.72& 1.52& 7.33& 1.48\\
           &   2M
& 0.19& 0.22& 0.17& 1.56& 3.35& 1.04& 0.76& 1.59& 6.30& 1.16\\
           &   4M
& 0.34& 0.38& 0.33& 1.55& 3.32& 1.03& 0.92& 1.91& 6.79& 0.92\\
           &   8M
& 0.66& 0.66& 0.65& 1.54& 2.91& 1.03& 1.01& 2.08& 7.48& 0.72\\
           &  16M
& 1.26& 1.27& 1.26& 1.53& 3.05& 1.02& 1.02& 2.11& 7.51& 0.57\\
m3killer   &  50K
& 0.01& 0.05& 0.01& 1.80& 5.05& 1.22& 0.46& 1.02& 7.91& 4.14\\
           & 100K
& 0.01& 0.05& 0.01& 1.74& 4.95& 1.16& 0.43& 0.96& 6.82& 3.19\\
           & 500K
& 0.05& 0.06& 0.05& 1.63& 4.22& 1.08& 0.61& 1.30& 7.76& 1.86\\
           &   1M
& 0.11& 0.11& 0.11& 1.60& 4.06& 1.06& 0.58& 1.23& 7.94& 1.46\\
           &   2M
& 0.17& 0.17& 0.16& 1.57& 3.76& 1.04& 0.69& 1.45& 8.19& 1.15\\
           &   4M
& 0.36& 0.39& 0.33& 1.56& 3.80& 1.03& 0.99& 2.04& 7.42& 0.92\\
           &   8M
& 0.69& 0.71& 0.66& 1.55& 3.60& 1.03& 0.94& 1.89& 8.60& 0.72\\
           &  16M
& 1.34& 1.38& 1.32& 1.54& 3.67& 1.02& 1.02& 2.05& 7.03& 0.57\\
\hline
\end{tabular}
\end{center}
\end{table}
The results for the sorted and rotated sequences are almost the same,
whereas the solution times on the organpipe and m3killer sequences
are between those for the sorted and random sequences.

The performance of {\sc pmSelect} on the same inputs is given in
Tables \ref{tab:pmSelrand} and \ref{tab:pmSeldet}.%
%
%   *** TABLE 7.3 ***
\begin{table}[t!]
\caption{Performance of {\sc pmSelect} on randomly generated inputs.}
\label{tab:pmSelrand}
\footnotesize
\begin{center}
\begin{tabular}{lrrrrrrrrrrrrr}
\hline
Sequence &\multicolumn{1}{c}{Size}
&\multicolumn{3}{c}{Time $[{\rm sec}]$%
\vphantom{$1^{2^3}$}} % Need more vertical space!
&\multicolumn{3}{c}{Comparisons $[n]$}
&\multicolumn{1}{c}{$\gamma_{\rm avg}$}
&\multicolumn{1}{c}{$L_{\rm avg}$}
&\multicolumn{1}{c}{$P_{\rm avg}$}
&\multicolumn{1}{c}{$N_{\rm avg}$}
&\multicolumn{1}{c}{$p_{\rm avg}$}
&\multicolumn{1}{c}{$s_{\rm avg}$}\\
&\multicolumn{1}{c}{$n$}
&\multicolumn{1}{c}{avg}&\multicolumn{1}{c}{max}&\multicolumn{1}{c}{min}
&\multicolumn{1}{c}{avg}&\multicolumn{1}{c}{max}&\multicolumn{1}{c}{min}
& &\multicolumn{1}{c}{$[n]$}
&\multicolumn{1}{c}{$[\ln n]$}
&\multicolumn{1}{c}{$[\ln n]$} &
&\multicolumn{1}{c}{$[\%n]$}\\
\hline
%alpha=0.5 beta=0.25 cutoff=600
random     &  50K
& 0.01& 0.06& 0.01& 1.79& 1.84& 1.74& 4.91& 1.21& 0.46& 1.01& 7.40& 4.10\\
           & 100K
& 0.01& 0.06& 0.01& 1.73& 1.77& 1.70& 4.77& 1.15& 0.43& 0.96& 8.03& 3.20\\
           & 500K
& 0.05& 0.06& 0.05& 1.62& 1.63& 1.61& 4.06& 1.08& 0.56& 1.20& 8.00& 1.86\\
           &   1M
& 0.11& 0.11& 0.11& 1.59& 1.60& 1.58& 3.95& 1.06& 0.67& 1.40& 7.95& 1.47\\
           &   2M
& 0.21& 0.22& 0.16& 1.57& 1.58& 1.56& 3.76& 1.04& 0.76& 1.59& 7.90& 1.16\\
           &   4M
& 0.39& 0.44& 0.38& 1.56& 1.56& 1.55& 3.63& 1.03& 0.95& 1.95& 7.29& 0.92\\
           &   8M
& 0.76& 0.77& 0.71& 1.54& 1.55& 1.54& 3.54& 1.03& 0.98& 2.00& 7.41& 0.72\\
           &  16M
& 1.49& 1.54& 1.48& 1.53& 1.54& 1.53& 3.39& 1.02& 1.00& 2.05& 7.77& 0.57\\
onezero    &  50K
& 0.01& 0.06& 0.01& 1.60& 1.60& 1.58& 1.64& 1.10& 0.46& 1.01& 5.63& 3.72\\
           & 100K
& 0.02& 0.06& 0.01& 1.58& 1.58& 1.56& 1.57& 1.08& 0.43& 0.95& 6.06& 2.94\\
           & 500K
& 0.05& 0.06& 0.01& 1.54& 1.55& 1.52& 1.46& 1.04& 0.66& 1.39& 5.98& 1.79\\
           &   1M
& 0.07& 0.11& 0.05& 1.54& 1.54& 1.54& 1.49& 1.04& 0.67& 1.42& 6.37& 1.42\\
           &   2M
& 0.19& 0.22& 0.16& 1.53& 1.53& 1.53& 1.47& 1.03& 0.83& 1.72& 6.28& 1.14\\
           &   4M
& 0.36& 0.39& 0.32& 1.52& 1.52& 1.52& 1.42& 1.02& 1.43& 2.92& 5.37& 0.92\\
           &   8M
& 0.71& 0.72& 0.71& 1.52& 1.52& 1.52& 1.40& 1.02& 1.53& 3.13& 5.54& 0.72\\
           &  16M
& 1.42& 1.43& 1.37& 1.51& 1.51& 1.51& 1.38& 1.01& 1.76& 3.58& 5.54& 0.58\\
twofaced   &  50K
& 0.01& 0.06& 0.01& 1.81& 1.84& 1.76& 5.11& 1.21& 0.46& 1.02& 7.78& 4.13\\
           & 100K
& 0.02& 0.06& 0.01& 1.73& 1.77& 1.71& 4.81& 1.16& 0.44& 0.96& 8.01& 3.20\\
           & 500K
& 0.06& 0.11& 0.05& 1.62& 1.63& 1.59& 4.10& 1.08& 0.56& 1.20& 8.15& 1.86\\
           &   1M
& 0.11& 0.11& 0.05& 1.59& 1.60& 1.58& 3.89& 1.06& 0.64& 1.36& 7.82& 1.47\\
           &   2M
& 0.20& 0.22& 0.16& 1.57& 1.58& 1.56& 3.63& 1.04& 0.75& 1.58& 7.63& 1.16\\
           &   4M
& 0.39& 0.44& 0.38& 1.56& 1.56& 1.55& 3.57& 1.03& 0.96& 1.97& 7.29& 0.92\\
           &   8M
& 0.76& 0.77& 0.71& 1.54& 1.55& 1.54& 3.50& 1.03& 0.97& 2.00& 7.43& 0.72\\
           &  16M
& 1.49& 1.54& 1.48& 1.53& 1.54& 1.53& 3.40& 1.02& 1.00& 2.03& 7.57& 0.57\\
\hline
\end{tabular}
\end{center}
\end{table}
%
%   *** TABLE 7.4 ***
\begin{table}[t!]
\caption{Performance of {\sc pmSelect} on deterministic inputs.}
\label{tab:pmSeldet}
\footnotesize
\begin{center}
\begin{tabular}{lrrrrcrrrrrr}
\hline
Sequence &\multicolumn{1}{c}{Size}
&\multicolumn{3}{c}{Time $[{\rm sec}]$%
\vphantom{$1^{2^3}$}} % Need more vertical space!
&\multicolumn{1}{c}{Comparisons}
&\multicolumn{1}{c}{$\gamma_{\rm avg}$}
&\multicolumn{1}{c}{$L_{\rm avg}$}
&\multicolumn{1}{c}{$P_{\rm avg}$}
&\multicolumn{1}{c}{$N_{\rm avg}$}
&\multicolumn{1}{c}{$p_{\rm avg}$}
&\multicolumn{1}{c}{$s_{\rm avg}$}\\
&\multicolumn{1}{c}{$n$}
&\multicolumn{1}{c}{avg}&\multicolumn{1}{c}{max}&\multicolumn{1}{c}{min}
&\multicolumn{1}{c}{$[n]$}
& &\multicolumn{1}{c}{$[n]$}
&\multicolumn{1}{c}{$[\ln n]$}
&\multicolumn{1}{c}{$[\ln n]$} &
&\multicolumn{1}{c}{$[\%n]$}\\
\hline
%alpha=0.5 beta=0.25 cutoff=600
sorted     &  50K
& 0.01& 0.06& 0.01& 1.79& 4.91& 1.23& 0.46& 1.02& 8.36& 4.12\\
           & 100K
& 0.03& 0.06& 0.01& 1.73& 4.69& 1.16& 0.43& 0.96& 8.55& 3.21\\
           & 500K
& 0.06& 0.06& 0.06& 1.60& 3.33& 1.07& 0.61& 1.30& 7.71& 1.86\\
           &   1M
& 0.07& 0.11& 0.05& 1.57& 3.07& 1.06& 0.65& 1.38& 6.58& 1.47\\
           &   2M
& 0.12& 0.17& 0.11& 1.56& 2.99& 1.04& 0.76& 1.59& 7.57& 1.15\\
           &   4M
& 0.25& 0.28& 0.22& 1.55& 3.02& 1.03& 0.99& 2.04& 8.06& 0.92\\
           &   8M
& 0.46& 0.49& 0.44& 1.54& 3.12& 1.03& 1.01& 2.01& 7.13& 0.72\\
           &  16M
& 0.90& 0.93& 0.88& 1.53& 3.20& 1.02& 1.02& 2.11& 7.46& 0.57\\
rotated    &  50K
& 0.01& 0.06& 0.01& 1.80& 4.92& 1.23& 0.46& 1.02& 8.55& 4.12\\
           & 100K
& 0.01& 0.06& 0.01& 1.73& 4.69& 1.16& 0.43& 0.96& 8.55& 3.21\\
           & 500K
& 0.03& 0.06& 0.01& 1.60& 3.33& 1.07& 0.61& 1.30& 7.82& 1.86\\
           &   1M
& 0.06& 0.11& 0.05& 1.57& 3.08& 1.06& 0.65& 1.38& 6.74& 1.47\\
           &   2M
& 0.12& 0.17& 0.11& 1.56& 2.99& 1.04& 0.76& 1.59& 7.13& 1.15\\
           &   4M
& 0.24& 0.28& 0.22& 1.55& 3.02& 1.03& 0.99& 2.04& 7.71& 0.92\\
           &   8M
& 0.46& 0.50& 0.44& 1.54& 3.12& 1.03& 1.01& 2.01& 7.19& 0.72\\
           &  16M
& 0.91& 0.93& 0.88& 1.53& 3.20& 1.02& 1.02& 2.11& 7.46& 0.57\\
organpipe  &  50K
& 0.02& 0.06& 0.01& 1.82& 5.26& 1.21& 0.46& 1.02& 8.73& 4.11\\
           & 100K
& 0.01& 0.01& 0.01& 1.76& 5.29& 1.17& 0.43& 0.96& 8.64& 3.20\\
           & 500K
& 0.06& 0.06& 0.06& 1.62& 3.95& 1.07& 0.61& 1.30& 7.06& 1.87\\
           &   1M
& 0.09& 0.11& 0.05& 1.59& 3.76& 1.06& 0.72& 1.52& 7.43& 1.48\\
           &   2M
& 0.16& 0.17& 0.16& 1.57& 3.37& 1.04& 0.76& 1.59& 7.00& 1.16\\
           &   4M
& 0.32& 0.33& 0.27& 1.55& 3.35& 1.03& 0.92& 1.91& 6.90& 0.92\\
           &   8M
& 0.56& 0.60& 0.55& 1.54& 2.91& 1.03& 1.01& 2.08& 7.97& 0.72\\
           &  16M
& 1.11& 1.15& 1.10& 1.53& 3.05& 1.02& 1.02& 2.11& 7.34& 0.57\\
m3killer   &  50K
& 0.01& 0.01& 0.01& 1.80& 5.05& 1.22& 0.46& 1.02& 7.91& 4.14\\
           & 100K
& 0.01& 0.05& 0.01& 1.74& 4.95& 1.16& 0.43& 0.96& 6.82& 3.19\\
           & 500K
& 0.05& 0.06& 0.05& 1.63& 4.22& 1.08& 0.61& 1.30& 7.76& 1.86\\
           &   1M
& 0.09& 0.11& 0.05& 1.60& 4.06& 1.06& 0.58& 1.23& 7.94& 1.46\\
           &   2M
& 0.17& 0.17& 0.16& 1.57& 3.76& 1.04& 0.69& 1.45& 8.19& 1.15\\
           &   4M
& 0.33& 0.33& 0.33& 1.56& 3.80& 1.03& 0.99& 2.04& 7.42& 0.92\\
           &   8M
& 0.61& 0.66& 0.60& 1.55& 3.60& 1.03& 0.94& 1.89& 8.60& 0.72\\
           &  16M
& 1.22& 1.26& 1.21& 1.54& 3.67& 1.02& 1.02& 2.05& 7.03& 0.57\\
\hline
\end{tabular}
\end{center}
\end{table}
{\sc Select} is slower than {\sc pmSelect} (but not too much: about
9\% on random and twofaced, 44\% on onezero, 16\% on sorted and rotated,
13\% on organpipe, 10\% on m3killer).  Except for timings and the onezero
results, Tables \ref{tab:pmSelrand}--\ref{tab:pmSeldet} almost coincide
with \ref{tab:Selrand}--\ref{tab:Seldet}.

The performance of {\sc riSelect} on the same inputs is described in
Tables \ref{tab:riSelrand} and \ref{tab:riSeldet}, where $N_{\rm rnd}$
denotes the average number of randomization steps.
%
%   *** TABLE 7.5 ***
\begin{table}[t!]
\caption{Performance of {\sc riSelect} on randomly generated inputs.}
\label{tab:riSelrand}
\footnotesize
\begin{center}
\begin{tabular}{lrrrrrrrrr}
\hline
Sequence &\multicolumn{1}{c}{Size}
&\multicolumn{3}{c}{Time $[{\rm sec}]$%
\vphantom{$1^{2^3}$}} % Need more vertical space!
&\multicolumn{3}{c}{Comparisons $[n]$}
&\multicolumn{1}{c}{$L_{\rm avg}$}
&\multicolumn{1}{c}{$N_{\rm rnd}$}\\
&\multicolumn{1}{c}{$n$}
&\multicolumn{1}{c}{avg}&\multicolumn{1}{c}{max}&\multicolumn{1}{c}{min}
&\multicolumn{1}{c}{avg}&\multicolumn{1}{c}{max}&\multicolumn{1}{c}{min}
&\multicolumn{1}{c}{$[n]$}&\\
\hline
random   &  50K & 0.01& 0.06& 0.01& 3.10& 4.32& 1.88&    3.10 & 0.40\\
         & 100K & 0.03& 0.06& 0.01& 2.61& 4.20& 1.77&    2.61 & 0.25\\
         & 500K & 0.10& 0.11& 0.05& 2.90& 4.23& 1.69&    2.90 & 0.20\\
         &   1M & 0.18& 0.22& 0.11& 2.81& 3.64& 1.84&    2.81 & 0.35\\
         &   2M & 0.34& 0.44& 0.22& 2.60& 3.57& 1.83&    2.60 & 0.30\\
         &   4M & 0.77& 1.38& 0.44& 2.88& 4.81& 1.83&    2.88 & 0.55\\
         &   8M & 1.38& 1.70& 1.05& 2.60& 3.48& 1.80&    2.60 & 0.45\\
         &  16M & 3.00& 4.01& 1.75& 2.99& 4.49& 1.73&    2.99 & 0.45\\
onezero  &  50K & 0.02& 0.06& 0.01& 2.73& 3.22& 2.68&    2.73 & 0.00\\
         & 100K & 0.03& 0.06& 0.01& 2.72& 2.88& 2.68&    2.72 & 0.00\\
         & 500K & 0.11& 0.17& 0.06& 2.74& 2.88& 2.68&    2.74 & 0.40\\
         &   1M & 0.20& 0.22& 0.16& 2.72& 2.85& 2.68&    2.72 & 0.55\\
         &   2M & 0.39& 0.44& 0.38& 2.71& 2.99& 2.68&    2.71 & 0.75\\
         &   4M & 0.79& 0.83& 0.76& 2.73& 2.85& 2.68&    2.73 & 1.00\\
         &   8M & 1.62& 1.98& 1.54& 2.72& 2.88& 2.68&    2.72 & 1.00\\
         &  16M & 3.13& 3.19& 3.07& 2.72& 2.85& 2.68&    2.72 & 0.95\\
twofaced &  50K & 0.03& 0.11& 0.01& 7.74& 8.45& 7.00&    7.74 & 1.20\\
         & 100K & 0.05& 0.11& 0.01& 7.57& 8.35& 6.79&    7.57 & 1.20\\
         & 500K & 0.17& 0.22& 0.11& 7.60& 9.25& 6.60&    7.60 & 1.25\\
         &   1M & 0.35& 0.39& 0.27& 7.64& 8.61& 7.02&    7.64 & 1.35\\
         &   2M & 0.70& 0.77& 0.55& 7.69& 8.55& 6.72&    7.69 & 1.30\\
         &   4M & 1.39& 1.65& 1.21& 7.70& 8.98& 6.89&    7.70 & 1.30\\
         &   8M & 2.80& 3.30& 2.47& 7.73& 9.12& 6.97&    7.73 & 1.30\\
         &  16M & 5.39& 6.15& 4.83& 7.49& 8.34& 6.79&    7.49 & 1.40\\
\hline
\end{tabular}
\end{center}
\end{table}
%
%   *** TABLE 7.6 ***
\begin{table}[t!]
\caption{Performance of {\sc riSelect} on deterministic inputs.}
\label{tab:riSeldet}
\footnotesize
\begin{center}
\begin{tabular}{lrrrrcrr}
\hline
Sequence &\multicolumn{1}{c}{Size}
&\multicolumn{3}{c}{Time $[{\rm sec}]$%
\vphantom{$1^{2^3}$}} % Need more vertical space!
&\multicolumn{1}{c}{Comparisons}
&\multicolumn{1}{c}{$L_{\rm avg}$}
&\multicolumn{1}{c}{$N_{\rm rnd}$}\\
&\multicolumn{1}{c}{$n$}
&\multicolumn{1}{c}{avg}&\multicolumn{1}{c}{max}&\multicolumn{1}{c}{min}
&\multicolumn{1}{c}{$[n]$}
&\multicolumn{1}{c}{$[n]$}
&\\
\hline
sorted   &  50K & 0.01& 0.01& 0.01& 1.00& 1.00 & 0.00\\
         & 100K & 0.01& 0.06& 0.01& 1.00& 1.00 & 0.00\\
         & 500K & 0.01& 0.01& 0.01& 1.00& 1.00 & 0.00\\
         &   1M & 0.05& 0.11& 0.01& 1.00& 1.00 & 0.00\\
         &   2M & 0.08& 0.11& 0.05& 1.00& 1.00 & 0.00\\
         &   4M & 0.15& 0.17& 0.11& 1.00& 1.00 & 0.00\\
         &   8M & 0.29& 0.33& 0.27& 1.00& 1.00 & 0.00\\
         &  16M & 0.56& 0.60& 0.55& 1.00& 1.00 & 0.00\\
rotated  &  50K & 0.01& 0.06& 0.01& 3.99& 3.98 & 2.00\\
         & 100K & 0.02& 0.06& 0.01& 3.97& 3.97 & 2.00\\
         & 500K & 0.11& 0.16& 0.06& 4.01& 4.01 & 3.00\\
         &   1M & 0.13& 0.17& 0.11& 3.96& 3.96 & 2.00\\
         &   2M & 0.28& 0.33& 0.27& 3.99& 3.99 & 1.00\\
         &   4M & 0.56& 0.60& 0.55& 4.00& 4.00 & 3.00\\
         &   8M & 1.10& 1.10& 1.10& 3.97& 3.97 & 2.00\\
         &  16M & 2.19& 2.20& 2.15& 3.96& 3.96 & 2.00\\
organpipe&  50K & 0.01& 0.06& 0.01& 9.43& 9.43 & 4.00\\
         & 100K & 0.06& 0.11& 0.01& 9.73& 9.73 & 4.00\\
         & 500K & 0.16& 0.17& 0.16& 8.31& 8.31 & 4.00\\
         &   1M & 0.35& 0.38& 0.33& 8.53& 8.53 & 5.00\\
         &   2M & 0.77& 0.77& 0.77& 9.73& 9.73 & 5.00\\
%        &   4M & 1.87& 1.87& 1.87&12.10&12.10 & 5.00\\
         &   4M & 1.87& 1.87& 1.87&\phantom{1.23}\llap{12.10}&12.10 & 5.00\\
         &   8M & 2.25& 2.26& 2.25& 7.34& 7.34 & 3.00\\
         &  16M & 5.07& 5.11& 5.05& 7.88& 7.88 & 3.00\\
m3killer &  50K & 0.02& 0.06& 0.01& 7.57& 7.57 & 2.00\\
%        & 100K & 0.03& 0.06& 0.01&11.52&11.52 & 2.00\\
         & 100K & 0.03& 0.06& 0.01&\phantom{1.23}\llap{11.52}&11.52 & 2.00\\
         & 500K & 0.16& 0.17& 0.16& 7.64& 7.64 & 1.00\\
         &   1M & 0.33& 0.33& 0.33& 8.00& 8.00 & 1.00\\
         &   2M & 0.68& 0.71& 0.66& 8.26& 8.26 & 1.00\\
         &   4M & 1.13& 1.16& 1.09& 7.15& 7.15 & 1.00\\
         &   8M & 2.86& 2.86& 2.85& 9.19& 9.19 & 2.00\\
         &  16M & 4.72& 4.73& 4.72& 7.43& 7.43 & 2.00\\
\hline
\end{tabular}
\end{center}
\end{table}
Note that for {\sc riSelect}, $C_{\rm avg}\approx L_{\rm avg}$, i.e.,
the cost of median-of-3 finding is negligible.  On the random
sequences, the expected value of $C_{\rm avg}$ is of order $2.75n$
\cite{kimapr:ahf}, but Table \ref{tab:riSelrand} exhibits significant
fluctuations in the numbers of comparisons made.  The results for the
onezero sequences confirm that quicksort-like partitioning may handle
equal keys quite efficiently \cite{sed:qek}.  The results for the
twofaced, rotated and m3killer inputs are quite good, since some
versions of quickselect may behave very poorly on these inputs
\cite{val:iss} (note that we used the ``sorted-median'' partitioning
variant as suggested in \cite{val:iss}).  Finally, the median-of-3
strategy employed by {\sc riSelect} really shines on the sorted inputs.

As always, limited testing doesn't warrant firm conclusions, but a
comparison of {\sc Select} and {\sc riSelect} is in order, especially
for the random sequences, which are most frequently used in theory and
practice for evaluating sorting and selection algorithms.  On the
random inputs, the ratio of the expected numbers of comparisons for
{\sc riSelect} and {\sc Select} is asymptotically $2.75/1.5\approx1.83$;
incidentally, the ratio of their computing times approaches
$3/1.62\approx1.85$ (cf.\ Tabs.\ \ref{tab:Selrand} and
\ref{tab:riSelrand}).  Note that {\sc Select} isn't just asymptotically
faster; in fact {\sc riSelect} is about 50\% slower even on middle-sized
inputs.  The same slow-down factor of about 50\% is observed on the
onezero sequences.  The performance gains of {\sc Select} over
{\sc riSelect} are much more pronounced on the remaining inputs, except
for the sorted sequences on which {\sc Select} may be slower by up to
88\%.  (However, the sorted input is quite special: increasing $k$ by
$1$ (for the upper median) doubled the solution times of {\sc riSelect}
without influencing those of {\sc Select} and {\sc pmSelect}; e.g., for
$n=16\mbox{\rm M}$ the respective times were $1.12$, $1.07$ and $0.90$).
Note that, relative to {\sc riSelect}, the solution times and comparison
counts of {\sc Select} and {\sc pmSelect} are much more stable across
all the inputs.  This feature may be important in applications.

%\cite{bad:irs,che:mae,dupa:cma,frmc:ssa}

%{\bf Acknowledgment}.  I would like to thank the Associate Editor and
%the two anonymous referees for their helpful comments.
{\bf Acknowledgment}.  I would like to thank Olgierd Hryniewicz,
Roger Koenker, Ronald L. Rivest and John D. Valois for useful
discussions.

%
%   *** REFERENCES ***
\footnotesize
%\bibliography{kckabbr,kalg,kbk,kck,kint,kth}
%\bibliographystyle{kck}
\newcommand{\etalchar}[1]{$^{#1}$}
\newcommand{\noopsort}[1]{} \newcommand{\printfirst}[2]{#1}
  \newcommand{\singleletter}[1]{#1} \newcommand{\switchargs}[2]{#2#1}
\ifx\undefined\bysame
\newcommand{\bysame}{\leavevmode\hbox to3em{\hrulefill}\,}
\fi

\normalsize
%   *** END OF REFERENCES ***
%
\end{document}